\documentclass[traditabstract]{aa}
\usepackage[]{hyperref}
\hypersetup{colorlinks=true,citecolor=blue}

\usepackage{color}

\usepackage{amsmath}
\usepackage{graphicx}
\usepackage{txfonts}
\usepackage{natbib}
\usepackage{url}

\usepackage[ruled]{algorithm2e}
\bibpunct{(}{)}{;}{a}{}{,}

\def\comment#1#2{}

\begin{document}

\title{Firedec: a two-channel finite-resolution image deconvolution algorithm}

\titlerunning{firedec: finite resolution deconvolution}

\author{N.~Cantale\inst{\ref{epfl}} \and F. Courbin\inst{\ref{epfl}}  \and M. Tewes\inst{\ref{bonn}, \ref{epfl}} \and P. Jablonka \inst{\ref{epfl}, \ref{paris}} \and G. Meylan\inst{\ref{epfl}}}

\institute{Laboratoire d'astrophysique, Ecole Polytechnique F\'ed\'erale de Lausanne (EPFL), Observatoire de Sauverny, CH-1290 Versoix, Switzerland, \email{frederic.courbin@epfl.ch} \label{epfl}
\and
Argelander-Institut f\"ur Astronomie, Auf dem H\"ugel 71, D-53121 Bonn, Germany \label{bonn}
\and
GEPI, Observatoire de Paris, CNRS UMR 8111, Universit\'e Paris Diderot, 92125, Meudon Cedex, France  \label{paris}
 }


\abstract{We present a two-channel deconvolution method  that decomposes images into a parametric point-source channel and a pixelized extended-source channel. Based on the central idea of the deconvolution algorithm proposed by Magain, Courbin \& Sohy (1998), the method aims at improving the resolution of the data rather than at
completely removing the point spread function (PSF). Improvements over the original method include a better regularization of the pixel channel of the image, based on wavelet filtering and multiscale analysis, and a better controlled separation of the point source vs. the extended source. In addition, the method is able to simultaneously deconvolve many individual frames of the same object taken with different instruments under different PSF conditions. For this purpose, we introduce a general geometric transformation between individual images. This transformation allows the combination of the images without having to interpolate them. We illustrate the capability of our algorithm using real and simulated images with complex diffraction-limited PSF. 
}

\maketitle

\section{Introduction}
\label{intro}

Imaging at high angular resolution can be achieved in three different ways. First, by launching space telescopes; this way is free of any disturbance from the Earth atmosphere.  For wavelengths that do not reach the ground because they are fully absorbed
in the atmosphere, space observatories are in fact the only viable option ($\gamma$-rays, X-rays, UV, thermal IR). Second,  complex optical instruments can be devised that minimize the effect of atmospheric turbulence. These optical systems are based on adaptive optics (AO) and involve a vibrating mirror placed in the optical path of the telescope. This distorts the observed wavefront in such a way that the distortion is opposite to the one produced by the air layers along the line of sight. AO is available on most major telescopes and will be essential for the future generation of giant ground-based telescopes such as the European Extremely Large Telescope (E-ELT) of the Thirty Meter Telescope (TMT). Finally, sharp images of the sky can also be reconstructed from interferograms obtained in the radio (VLA, IRAM, ALMA) or in the optical (e.g., VLTI). These require combining as many baselines as possible to cover the Fourier $(u, v)$ plane in the most efficient way. Giant interferometers such as the planned Square Kilometer Array (SKA) consider thousands of antennas with baselines in the range 1-100 km.

Improving these already excellent data further is always an advantage. Numerical techniques are therefore developed to postprocess the data: all instruments, even those located above the Earth atmosphere, produce images that are  blurred by the instrumental point spread function (PSF). When the observations are made from the ground, this blurring is increased by the additional atmospheric turbulence. Mathematically, blurring corresponds to a convolution by the PSF. When the data are sampled and affected by noise, inverting this operation is difficult. Image deconvolution can indeed become an ill-posed problem as soon as the convolution kernel is non-vanishing, that is, when it contains infinitely high spatial frequencies. This is most often the case in pixelized astronomical images.

Numerous methods have been proposed to deconvolve images. Among the most popular are Wiener filtering, the CLEAN algorithm, which
was first proposed by \citet{Hogbom1974} and is ex\-ten\-sively used in ra\-dio as\-tro\-no\-my, the maxi\-mum entropy method \citep[MEM; e.g.,][]{Skilling1984}, and the Richardson-Lucy algorithm \citep[RL;][]{Lucy1974, Richardson1972}, which became famous 
in the period during which the Hubble Space Telescope suffered from severe optical aberrations.

Other approaches have been adopted more recently, using direct filtering on multifrequency images \citep{Herranz2009} or by using independent component analysis (ICA) in combination with neural networks. \citet{Baccigalupi2000} started in this field, and \citet{Aumont2007} followed, with an algorithm called PolEMICA
that was later refined and is known as MILCA \citep{Hurier2013}. More recently, \citet{Bobin2013} developed the generalized multicomponent analysis (L-GMCA) algorithm, which is based on multiscale analysis and sparsity. \citet[][]{Lefkimmiatis2013} developed a deconvolution method in the context of Poisson noise using Hessian Schatten-norm regularization.

In addition, the application of source separation techniques to deconvolution algorithms has been proposed since \citet{Hook1994a}. These methods are known as two-channel methods and decompose the images into a point-source channel plus an extended-source channel. The RL algorithm has been modi\-fied to include this two-channel decomposition and is known as the PLUCY and then CPLUCY algorithm. This led to the GIRA task of the IRAF
software \citep{Pirzkal2000}. 

\citet{Velusamy2008} also proposed a second channel for point sources. Multichannel deconvolution methods were also applied to maximum entropy methods \citep[e.g.,][]{Bontekoe1993, Weir1992} and to the specific problem of deconvolving radio interferometric data \cite[e.g.,][]{Giovannelli2005}. The most recent development of a two-channel deconvolution method, to our knowledge, has
been proposed by \citet{Selig2015}. The method is derived in a Bayesian framework and exploits prior information on the spatial correlation structure of the extended component and the brightness distribution of the spatially uncorrelated point-like sources.

All the above methods have advantages and drawbacks and have been implemented in many different forms. However, the vast majority of them attempt to perfectly correct for the PSF, resulting in signi\-ficant artifacts that are due to Gibbs oscillations of
nearby sharp structures. Most of the work with improved versions of the above methods focuses on removing or minimizing these artifacts. In some cases, it is often tempting to stop the deconvolution process before convergence is reached to avoid these artifacts from becoming too prominent. 

We here describe a two-channel image deconvolution algorithm based on the central idea of the MCS algorithm \citep{MCS}, which deconvolves images using a PSF that is narrower than the observed one. This results in deconvolved images that are much less affected by artifacts. We extend the capabilities of the algorithm, notably by introducing an improved separation of the point source vs. the extended source and a modified regularization term that is
based on wavelet filtering. Like the original MCS algorithm, our new method preserves flux and astrometry and can process multiple exposures simultaneously.

The paper is organized as follows. In Sect. \ref{MCS} we recall the MCS deconvolution and give an overview of the algorithm. The problem of regularizing the pixel channel is presented in Sect. \ref{wavelet}. We describe in Sect. 4 the tool that we adopted and developed to address this problem; it is based on image de-noising by wavelet filtering. This same tool is also employed to improve the splitting of images into the parametric and pixelized channels (Sect. \ref{separation}) and to characterize the noise in the input images (Appendix \ref{noise}). In Sect. \ref{handbook} we return to the overall algorithm by summarizing how the optimization is performed. We present tests and demonstrations of the algorithm in Sect. \ref{tests} and conclude in Sect. \ref{conclusion}.

\begin{figure*}[t!]
\centering
\includegraphics[width=17cm]{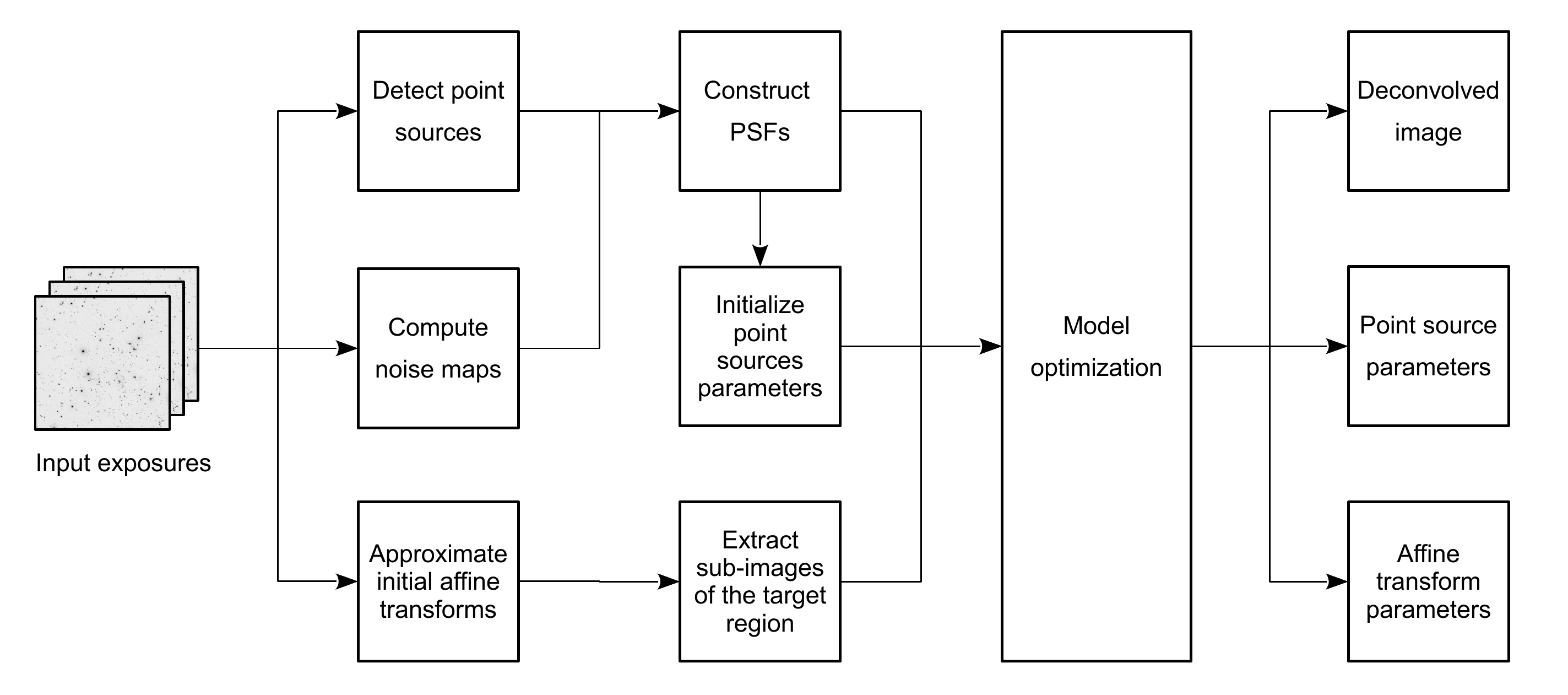}
\caption{Our algorithm uses a multi-exposure data set to set up the upstream parameters before proceeding with the optimization, which is detailed in Sect.~\ref{alg:minimi} and Fig.~\ref{fig:minimi_flow}. The deconvolution products are the deconvolved pixel image (decomposed into two channels), the point source parameters (fluxes and positions), and the geometric affine transform parameters for each input image.}
\label{fig:dec_flow}
\end{figure*}

\section{MCS deconvolution algorithm}
\label{MCS}

To simplify notations, all equations in this paper are given for one-dimensional data, but a generalization to two dimensions is straightforward. Variables in boldface are arrays of pixels. Operators and arithmetics in general apply pixel-wise to these arrays. For example, $\mathbf{A} = \mathbf{B}/\mathbf{C}$ means that all pixels of image $\mathbf{B}$ are divided by the pixels at the same coordinates in image $\mathbf{C}$ and stored in image $\mathbf{A}$. The asterisk denotes the convolution operator. We express all pixel intensities (fluxes) in units of electrons. 

A deconvolution algorithm attempts to undo what the con\-volu\-tion operation does, that is,
\begin{equation}
\mathbf{D} = (\mathbf{PSF} * \mathbf{M}) + \mathbf{Z},
\label{eqn:deconv}
\end{equation} 
where the data $\mathbf{D}$ can be represented as the convolution of a model $\mathbf{M}$ by the total $\mathbf{PSF}$ that is further affected by noise $\mathbf{Z}$. In astronomical images, this noise is often dominated by a Poisson-distributed term, but it
is usually approximated by Gaussian noise with a standard deviation that might vary from pixel to pixel.

Deconvolution, that is, estimating the model $\mathbf{M}$ given the data and the PSF, is difficult: many different models may be compatible with the data, given the noise. One way to address this is to minimize a cost function that includes a regularization term, 
\begin{equation}
C(\mathbf{M}) = \sum_{i=1,n} \left[ \dfrac{\mathbf{D} - (\mathbf{PSF*M})}{\boldsymbol{\sigma}}\right]_i^2 + \lambda\cdot H,
\label{eqn:cost}
\end{equation}
where the index, $i$, indicates the pixel number ($n$ being the total number) and where $\sigma_i$ is the rms noise at each pixel. The term $H$ is a  regularization weighted by a Lagrange parameter, $\lambda$. In practice, $H$ is often chosen as a Tikhonov regula\-rization.

\subsection{Target PSF}

The regularization term helps minimizing the parameter degeneracies and reducing the so-called deconvolution artifacts that are often seen as rings around structures with high spatial frequency. These rings are Gibbs oscillations, which are due to the finite sampling of the pixel grid used to represent the deconvolved image. 

The solution proposed in the MCS algorithm \citep{MCS} is to circumvent the problem of deconvolution artifacts by using  a kernel in
the deconvolution process  that is narrower than the observed PSF. Deconvolution performed with such a narrower PSF avoids violating the sampling theorem and allows properly sampling the structures with the highest spatial frequencies allowed in the deconvolved image. The desired PSF in the deconvolved image can
be chosen to be of any arbitrary shape as long as it can be at least critically sampled. An obvious choice is a circular Gaussian function that we call the {\it \textup{target}} PSF. The \textup{\textup{{\it \textup{observed}}}} PSF and the target PSF are related by the simple relation
\begin{equation}
\mathbf{PSF} = \mathbf{g} * \mathbf{P},
\label{eqn:psf_mcs}
\end{equation} 
where $\mathbf{P}$ is the kernel that transforms the observed PSF, $\mathbf{PSF}$, into the target PSF, $\mathbf{g}$. If we use the kernel $\mathbf{P}$ instead of $\mathbf{PSF}$ in  Eq.~(\ref{eqn:cost}), the target PSF in the deconvolved image is $\mathbf{g}$. As $\mathbf{g}$ is chosen to be properly sampled, Gibbs oscilations are minimized. This is the main idea behind the MCS algorithm, which aims at improving the resolution of the image, without attempting to achieve an infinite resolution.

\subsection{Two-channel deconvolved model}

The model image $\mathbf{M}$ can be decomposed into a sum of Gaussian point sources representing unresolved objects such as stars, and of a pixelized channel representing resolved sources, $\mathbf{B}$: 
\begin{equation}
\mathbf{M} (\mathbf{B},\vec{c},\vec{a}) = \mathbf{B} + \sum_{j=1,n_\star} a_{j}\ \mathbf{g}(x-c_{j}),
\end{equation}
which contains $n_\star$ point sources, each centered at position $c_j$ and with a total flux $a_j$. The parameters to vary when minimizing the cost function are therefore all the pixels contained in the pixel-channel image $\mathbf{B}$ and the positions and fluxes of all point sources contained in the parametric channel. 

\subsection{Sampling of the deconvolved image}

The main principle behind the MCS algorithm is to avoid violating the sampling theorem at any stage of the deconvolution process. What matters in this case is the sampling in the deconvolved image, even if the sampling of the original data is not optimal. For this reason the MCS algorithm uses a subsampled version of the deconvoved image, where the target resolution can always be properly sampled. This is mathematically defined with two sampling operators. In one dimension and with a factor of 2, up-sampling is given by
\begin{equation}
y(i) = x_i^\uparrow \triangleq 
\left\lbrace
\begin{array}{cc}
x\left(\frac{i}{2}\right)/2, & \qquad\frac{i}{2} \in \mathbb{N} \\
x\left(\frac{i-1}{2}\right)/2, & \qquad\frac{i}{2} \notin \mathbb{N} \\
\end{array}\right.,\\
\end{equation}
where $^\uparrow$ denotes the up-sampling operator. Likewise, the down-sampling operation takes the sum of adjacent pixels:
\begin{equation}
y(i) = x_i^\downarrow \triangleq x(2i)+x(2i+1).
\end{equation}

We note that following these definitions, the surface brightness of fixed sky area is the same before and after re-sampling. This would not be the case with a traditional operator taking the mean value in up-sampling and an equal value when down-sampling. The generalization to the two-dimensions case is trivial: the up-sampling is applied likewise on both coordinates and the down-sampling is the sum of all subpixels. Resampling to higher factors is achieved by successively applying these operators. In the rest of the paper, we use this definition of the subsampling operators.

\subsection{Cost function for multiple frames}

With all the above definitions and principles, the MCS image deconvolution, as implemented in its original form, minimizes the following cost function:
\begin{equation}
C_{\chi^2}  = \frac{1}{N} \sum_{i=1,n} \sum_{j=1,N} {\left [
\frac{\mathbf{D}_j - (\mathbf{P}_j*\mathbf{M})^\downarrow}
{\boldsymbol{\sigma}_j} \right ]_i^\uparrow}^2 ,
\label{eqn:cost_multi}
\end{equation}
where the $i$-index runs along the $n$ pixels of each of the $N$ images to be deconvolved simultaneously. This simultaneous deconvolution leads to a final deep and sharp combined frame, similar to the output of the \texttt{drizzle} algorithm \citep{drizzle}, but deconvolved and decomposed into two channels. 

%
In the original implementation of the algorithm, only shifts between the data frames were allowed. We now implement an affine transformation
between the model image $\mathbf{M}$ and each data frame $\mathbf{D}_j$,  
\begin{equation}
\begin{pmatrix}
   x^\prime \\
   y^\prime \\
\end{pmatrix} = 
\begin{pmatrix}
   a_{j,11} & a_{j,12} \\
   a_{j,21} & a_{j,22} \\
\end{pmatrix} \cdot 
\begin{pmatrix}
   x \\
   y \\
\end{pmatrix} + 
\begin{pmatrix}
   dx_j \\
   dy_j \\
\end{pmatrix} ,
\label{eq:afftrans}
\end{equation}
where the parameters of the transformation are $a_{j,lm}$ , which encode scaling, rotation, and shear, and $dx_j$ and $dy_j$ ,
which correspond to translations between the images. The new pixel flux at position $(x^\prime, y^\prime)$ is renormalized to conserve the surface brightness on the plane of the sky. In addition, we allow for an additional multiplicative factor, $t_j$, to account for images in different flux units, for example, different exposure times.

A potent consequence of this configuration is that each data image may now have its own sampling, orientation, deformation, or exposure time and may still be combined into a single deconvolved frame. Moreover, if the parameters of the transformation are not precisely known, they may be optimized during the deconvolution process.

If only dithered images with the same exposure time are considered, with no rotation and no distortion, then 
\begin{equation} 
a_{j,11}=a_{j,22}=t_j=1,\qquad a_{j,12}=a_{j,21}=0,
\end{equation}
and only the dithering parameters, $dx_j$ and $dy_j$, are non-zero. If each data frame is rotated by $\theta_j$ , then 
\begin{eqnarray}
a_{j,11} & = & +{\rm cos}(\theta_j), \nonumber \\ 
a_{j,12} & = & -{\rm sin}(\theta_j),\nonumber \\
a_{j,21} & = & +{\rm sin}(\theta_j), \nonumber \\
a_{j,22} & = & +{\rm cos}(\theta_j).
\end{eqnarray}

More general distortions, like shearing, can be represented with any combination of the $a_{j,lm}$ parameters.

A general overview of the deconvolution algorithm is given in Fig.~\ref{fig:dec_flow}. The different steps before the model optimization are displayed together with the deconvolution products. Before addressing the optimization problem itself (Sect.~\ref{handbook}), we detail in the next sections how we built our algorithm.

\section{Regularization of the pixel channel}
\label{wavelet}

The purpose of the regularization term, $\lambda H$, in Eqs.~\ref{eqn:cost} and \ref{eqn:cost_multi} is to penalize the high frequencies arising in the pixel channel from noise enhancement and Gibbs oscillations. In the MCS algorithm, Gibbs oscillations are suppressed or at least minimized, but noise enhancement still needs to be
considered.

A popular way to implement the regularization is to high-pass filter the deconvolved image by subtracting a smoothed (low-pass filtered) version of it. In the original MCS algorithm, the regularization term, applied only to the pixel channel $\mathbf{B}$ of the deconvolved image, is 
\begin{equation}
H_{\mathrm{MCS}} = \sum_{i=1,n} [\mathbf{B} - (\mathbf{B} * \mathbf{g})]_i^2,
\label{eqn:reg_MCS}
\end{equation}
where $\mathbf{g}$ is again the target PSF. This regularization term is multiplied by a Lagrange parameter $\lambda$, allowing us to weight its effect on the total cost function. However, the scaling between the $\chi^2$ term and the regularization is not linear and highly depends on the signal-to-noise ratio $S/N$, which can vary strongly across the image. This is true for any image, but even more so for the extreme dynamical range of astronomical images. To remedy this, we modified the regularization of the MCS algorithm so that it scales with $S/N$, that is, we ensured that high $S/N$ regions are treated in the same way as lower $S/N$ regions, without any need to implement a spatially variable weighting term $\lambda_i$. Our cost function takes the form
\begin{equation}
C_{\mathrm{pix}} = \frac{1}{N} \sum_{i=1,n} \sum_{j=1,N} {\left [
\frac{\mathbf{D}_j - (\mathbf{P}_j*\mathbf{M})^\downarrow}
{\boldsymbol{\sigma}_j} \right ]_i^\uparrow}^2  
+ \lambda \sum_{i=1,n}\left( \frac{\mathbf{B}-\tilde{\mathbf{B}}}{\sqrt{1+\mathbf{B}}} \right)_i^2,
\label{eqn:final_bkg} 
\end{equation}
which we rewrite in a more compact form as 
\begin{equation}
C_{\mathrm{pix}} = C_{\chi^2} + \lambda\cdot H_{\mathrm{reg}} .
\end{equation}
In these equations,  $\tilde{\mathbf{B}}$ is a de-noised version of $\mathbf{B}$:
\begin{equation}
\tilde{\mathbf{B}} = \phi(\mathbf{B}),
\label{phi}
\end{equation}
where the function $\phi$ is a low-pass filter or any de-noising filter (Sect.~\ref{denoising}). For instance, the Gaussian filter used in MCS (Eq.~\ref{eqn:reg_MCS}) is a possible solution for the de-noising function. This solution is motivated by the fact that spatial frequencies higher than those allowed by the target PSF $\mathbf{g}$ must not be present in the pixel channel. However, this smoothing also penalizes variability at lower spatial frequencies and in general tends to make the deconvolved image of lower resolution than it should. It also directly influences the shape of objects with different S/N. For these reasons, we replaced it with a regularization term that preserves both the shape and the flux while completely removing the frequencies below the sampling limit.

\begin{figure}[t!]
\includegraphics[width=8.9cm]{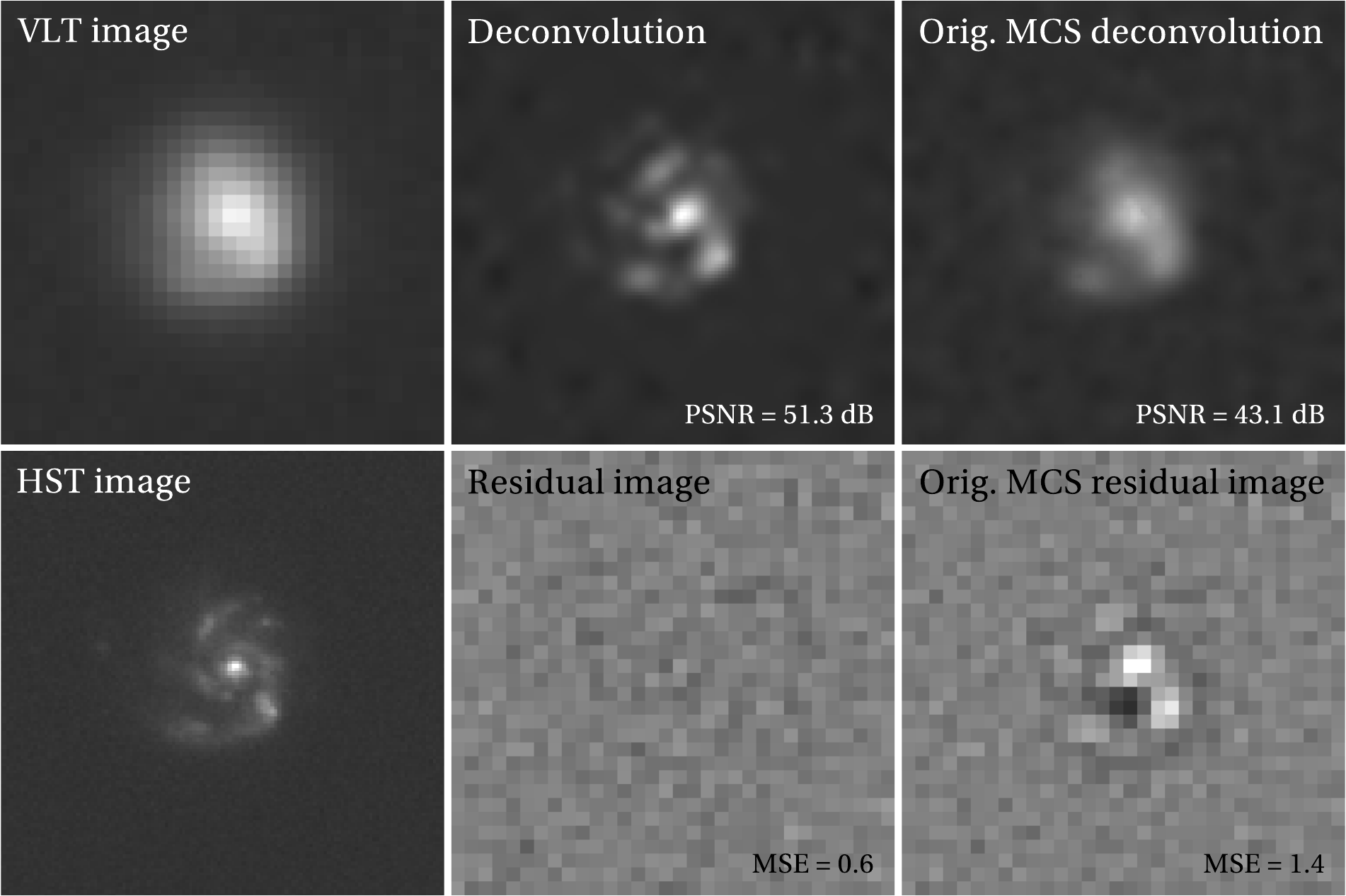}\caption{Effect of the regularization scheme on the deconvolution results. {\it Left:} a VLT/FORS2 optical image of a spiral galaxy is shown in the top panel, along with the HST/ACS observation of the same galaxy in the bottom panel. {\it Middle:} deconvolution with the new regularization scheme along with the residual image in the bottom panel. {\it Right:} same as in the middle panel, but with the regularization term used in the original MCS algorithm. The peak S/N (PSNR) is given for each deconvolution, and the mean square errors (MSE) are given with the residual images.}
\label{fig:reg_compar}
\end{figure}

\begin{figure*}[t!]
\centering
\includegraphics[width=14cm]{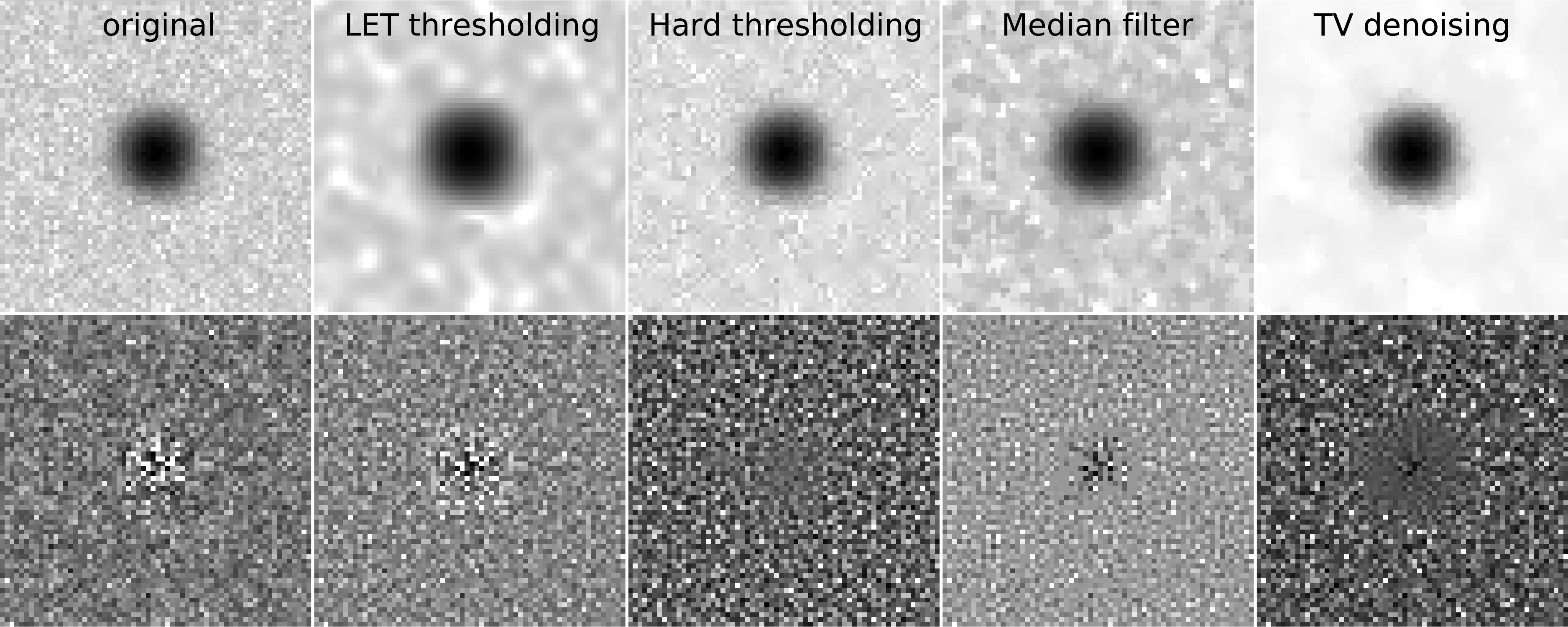}
\caption[Comparison of different denoising strategies]{Comparison of different de-noising strategies. {\it Top:} a simulated compact galaxy affected by Poisson noise, followed by different de-noised versions. {\it Bottom:} the noise image that has been removed by each de-noising scheme. The bottom left panel shows the original Poisson noise added to the simulated image. Clearly, the noise image recovered using a method designed for Poisson noise, i.e., in this case the LET thresholding (see text), is the closest to the original noise.}
\label{fig:dn_compar}
\end{figure*}

We illustrate in Fig.~\ref{fig:reg_compar} the effect of the modified regularization on the quality of the deconvolution. With the original MCS regularization, it is hardly possible to obtain good residuals both in the sky background and where an object is present. This is due to the dependence of the regularization on the local S/N in the image. In the present example, obtaining a good mean square error (MSE) in the center of the frame is possible only at the price of overfitting the noise in the outter parts. With the regularization based on wavelet de-noising, not only does this effect disappears, but the peak S/N (PSNR) improves as well, hence leading to a deconvolved image displaying a larger dynamical range.

\section{Image de-noising scheme}
\label{denoising}

The modified version of the MCS algorithm described in this paper makes use of image de-noising. This applies to the regularization term in Eq.~\ref{eqn:final_bkg} and to the point-source segregation scheme described in Sect.~\ref{separation}. In addition, we describe in Appendix~\ref{noise} how we use it to estimate the Poisson noise in the images to be deconvolved.

A critical part of a de-noising function is that it must preserve the shape as well as the flux of the images. Numerous methods have been proposed to remove noise from images. Some of the most popular ones involve a simple median filtering of the data or total variation de-noising \citep[TV-de-noising;][]{Rudin1992}. Other more sophisticated methods consider wavelet decomposition of the data followed by a thresholding scheme of the wavelet coefficients. 
 
A common way of de-noising using wavelets is to use the decimated Haar transform. In addition to being fast, this solution preserves the flux by construction: the local average flux in some location of the image is set by the high-level coefficients (low frequency), and the lower level coefficients (high frequency) represent the deviations from this local average value. This means that thresholding the low-level coefficients does not modify the mean local flux. However, as the decimated Haar transform is not shift-invariant, the de-noising result depends on where the original continuous signal is sampled on the pixel grid. The solution for circumventing this problem is to make use of cycle-spinning \citep{Coifman1995}. Cycle-spinning applies the Haar transform on shifted versions of the image, creating redundant versions of the image for all possible translations for which the de-noising is not invariant. After thresholding, the image is reconstructed by averaging over all these shifted versions. We combine the cycle-spinning with a modified Haar filter, the un-decimated bi-orthogonal Haar transform \citep{Zhang}. This filter has a larger kernel and produces more continuous results than the standard Haar decomposition. The simultaneous use of cycle-spinning and bi-Haar filters does not lead to artifacts that frequently occur when using decimated transforms.
We note that there are many alternatives to the above de-noising scheme. Examples can be found in \citet{starckbook2007} \& \citet{starckbook2010}, and they can be adapted to the specific type of spatial structures in the images to be deconvolved.

The choice of the thresholding strategy is also critical. While a standard ``hard thresholding'' works well for Gaussian noise, it does not perform well in the Poisson regime. Instead, we used a thresholding scheme inspired by \cite{Luisier2011}. The idea is to create a linear expansion of thresholds (LET) to adapt the threshold to the intensity level. This allows us to take into account Poisson noise, where the threshold is signal-dependent. For each wavelet coefficient, the threshold is estimated by considering the corresponding higher level coefficients, thus giving a good guess of the local signal level. 
For the $j$-th coefficient of the $i$-th level and for the image reference threshold $t$, the threshold is given by
\begin{equation}
t_{ij} = \sqrt{2^{-i/2}\cdot w_{i+1,j} +t^2} ,
\end{equation}
where $w_{i+1,j}$ is the low-frequency coefficient. The de-noised high-frequency ($i>0$) wavelet coefficients are then given by
\begin{equation}
\tilde{w}_{i,j} = w_{i,j} \cdot \left(1 - e^{-\left(w_{i,j} / 3t_{ij}\right)^8}\right) \;.
\end{equation}

This combination of a bi-Haar wavelet transform, cycle-spinning, and LET is well suited to our purpose, as illustrated in Fig.~\ref{fig:dn_compar}. We describe our de-noising function as
\begin{equation}
\phi = \phi_{n,t},
\end{equation}
where $n$ is the number of wavelet levels on which the threshold is applied and where $t$ is the threshold. In Eq.~\ref{eqn:final_bkg} the regularized version of the pixel channel of the image, $\mathbf{B}$, takes the form
\begin{equation}
\tilde{\mathbf{B}} = \phi_{1,\infty}(\mathbf{B}),
\end{equation}
meaning that the de-noising threshold of the pixel-channel image, $\mathbf{B}$, is set to infinity for the first level of coefficients. This configuration completely removes frequencies higher than one pixel, while the translation invariance (cycle-spinning) preserves the continuity of the image. In other words, any detail finer than two pixels is perceived as discontinuous and is regularized.

Note that our solution is only one option among many other possible choices. We deliberately chose the bi-Haar wavelet transform together with cycle-spinning and LET because this is  simple
to implement, but the operator $\phi$ in Eq.~\ref{phi} is generic and can well be replaced by any other method dealing with Poisson noise.

\section{Point-source separation}
\label{separation}

\comment{Malte}{ok with segregation instead of separation ? I rearranged a bit the sentences in this Section 5, check if ok.}

We now have a cost function 
with a regularization term that minimizes noise amplification. In minimizing the cost function, degeneracies inherent to the two-channel decomposition of the deconvolved image arise. Indeed, an infinite number of decompositions are possible between the pixel channel, $\mathbf{B}$, and the parametric channel, at least locally, in the immediate vicinity (1-2 pixels) of point sources. 

To circumvent the problem, we introduced 
an additional term to the overal cost function that penalizes gradients of the ave\-rage residual image that are defined as
\begin{equation}
\mathbf{R} = \frac{1}{N} \sum_{j=1,N} {\left (\mathbf{D}_j - \left(\mathbf{P}_j*\mathbf{M}\right)^\downarrow \right )^\uparrow}
\label{eq:res_deri}
,\end{equation}
where the summation is over the $N$ data frames considered in the calculation. As taking derivatives of a noisy image may increase the noise further, we considered a de-noised version $\tilde{\mathbf{R}} = \phi_{n,\infty}(\mathbf{R})$ of the residuals.\comment{malte}{Question : caption of Fig 5 uses only $n=1$, should we keep $\phi_{n,\infty}$ in the above sentence to stay general ?} We defined the associated contribution to the cost function using the norm of the gradients of $\tilde{\mathbf{R}}$ :
\begin{equation}
C_{\mathrm{param}} =  \left\| \frac{\partial^m\tilde{\mathbf{R}}}{ \partial x^m} \right\|^2,
\label{eq:deri_TV}
\end{equation}
where the exponent $m$ indicates the derivative order and $x$ is {\bf the  pixel} position in the image. We recall that for simplicity this equation is written here in its one-dimensional form. Depending on the type of image to be deconvolved, the derivative order $m$ can potentially be adapted after visual inspection of the $\chi^2$ images returned by the algorithm. In practice, taking second-order derivatives is often a good compromise between de-noising and the level of structures authorized in the pixel-channel image, $\mathbf{B}$.

\begin{figure}[t!]
\includegraphics[width=9.2cm]{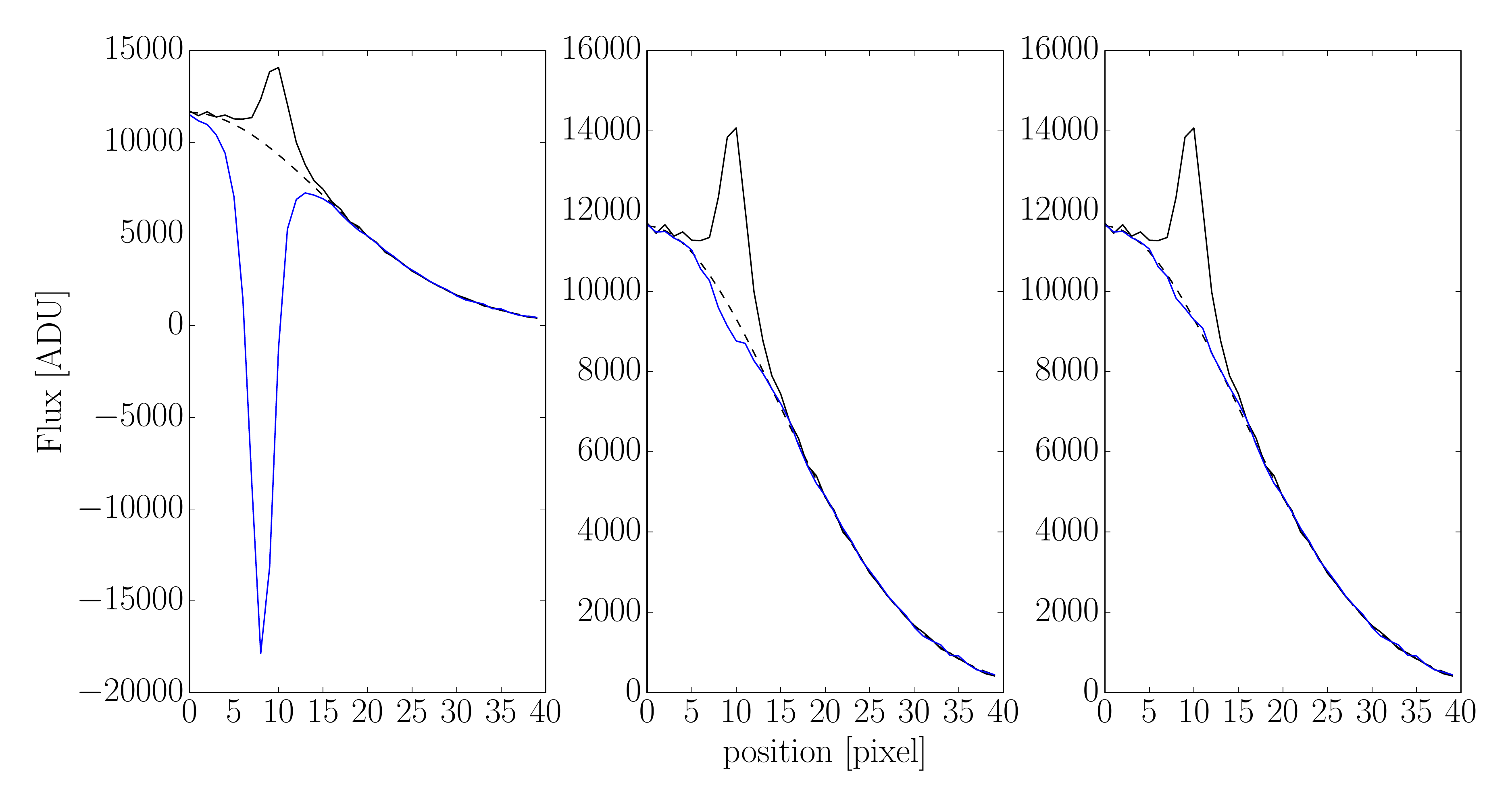}
\caption[]{Effect of the constraint on the residual derivatives (Eq.~\ref{eq:deri_TV}) on the separation of the point source and pixel channel. A radial cut through a simulated image is shown. {\it Left:} the black line shows the data, which is the sum of a point source and an extended source (dashed line). The deconvolution is performed with no constraint on the derivative image, resulting in a ``hole'' in the profile of the deconvolved extended source, shown in blue. {\it Middle:} the deconvolution is now made including a penalization of the first derivatives in the residuals. {\it Right:} same as middle, but using the second derivatives. The pixel channel is now well compatible with the input of the simulation. 
}
\label{fig:sep_prof}
\end{figure}

\begin{figure}[t!]
\centering
\includegraphics[width=8.8cm]{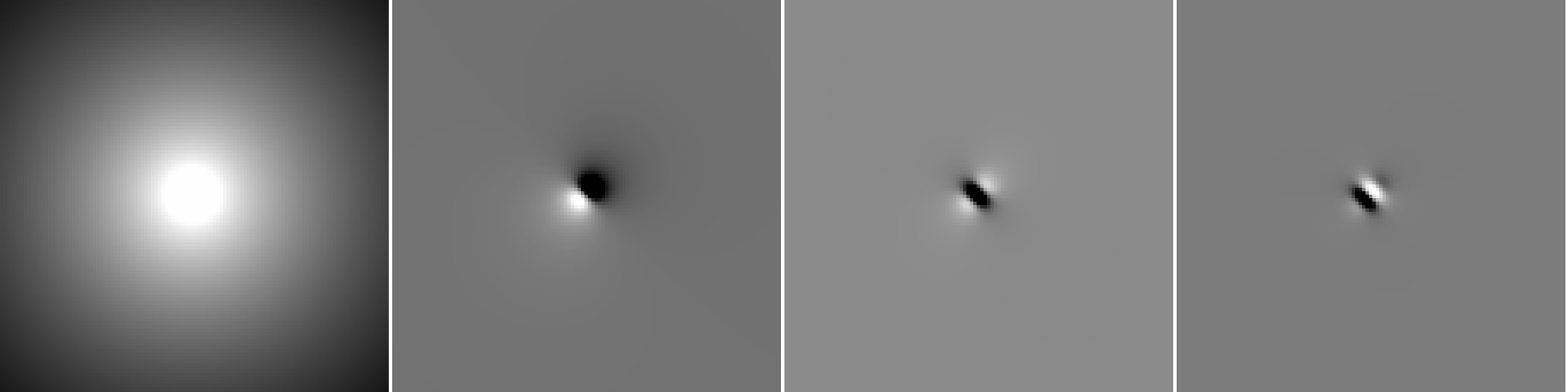}
\includegraphics[width=8.8cm]{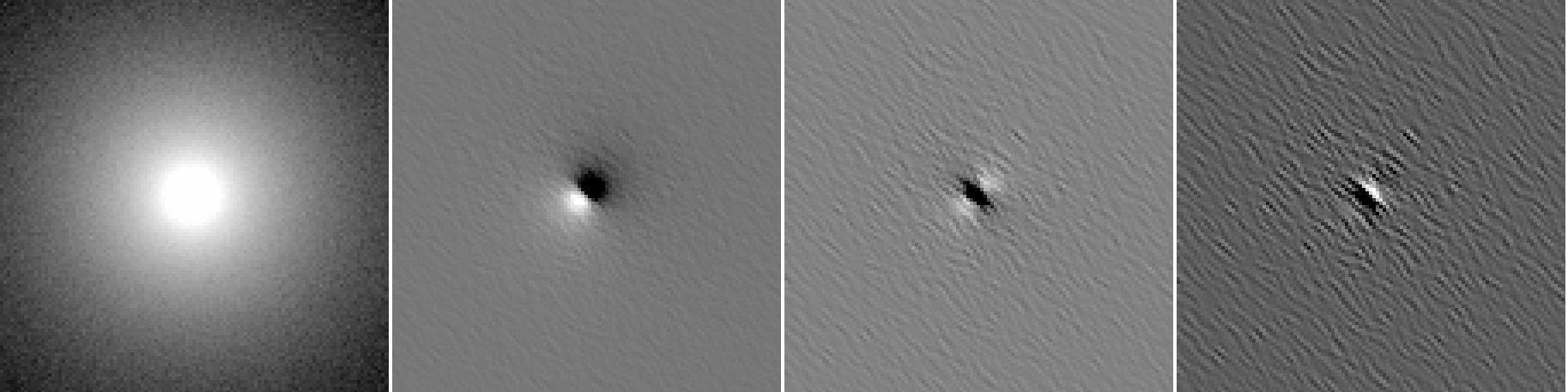}
\includegraphics[width=8.8cm]{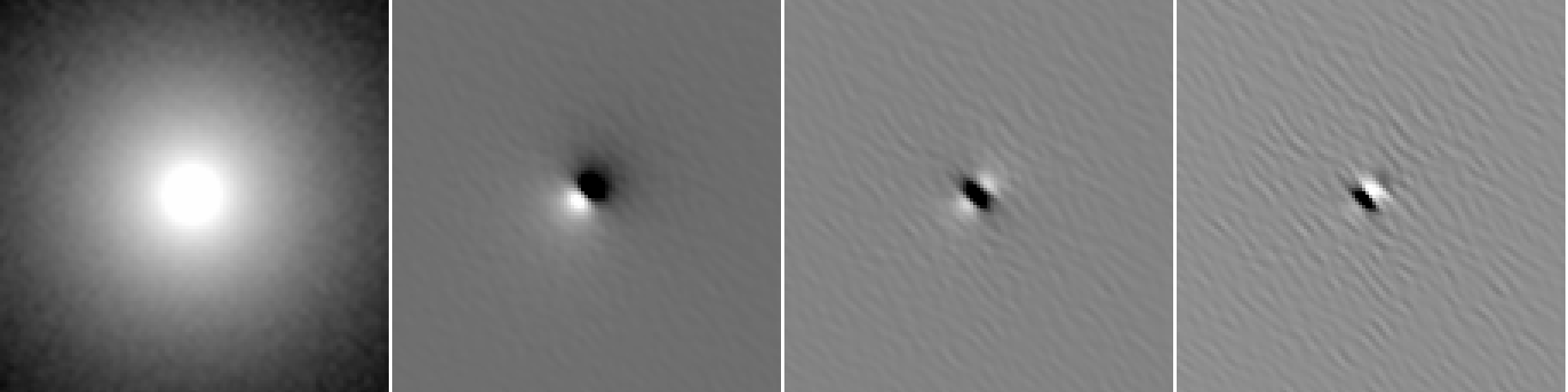}
\caption[]{Illustration of the effect of de-noising $\mathbf{R}$ on the derivatives of the residual image. From left to right we show a simulated image of a galaxy (Sersic profile) and the three first derivatives of the residual image $\mathbf{R}$ in Eq.~\ref{eq:res_deri}. The first row shows a noise-free image. The second row shows the derivatives of the non-de-noised residuals. The third row show the derivatives computed using $\tilde{\mathbf{R}} = \phi_{1,\infty}(\mathbf{R})$, with a significant improvement of the third derivatives. 
}
\label{fig:deriv_dn}
\end{figure}

The effect of this additional cost function term on the separation
of the point source and the extended source  is shown in Fig.~\ref{fig:sep_prof}. This term is crucial in obtaining a physically plausible solution to the separation. Figure \ref{fig:deriv_dn} illustrates the effect of denoising $\mathbf{R}$. We note that for high S/N data, $\mathbf{R}$ could be used in Eq. \ref{eq:deri_TV} directly without consequences, but it is useful to use the de-noised image, $\tilde{\mathbf{R}}$, for low S/N data.


%

The final deconvolution problem is thus described by optimizing the following objective function:
\begin{equation}
C =  C_{\chi^2} + \lambda\cdot H_{\mathrm{reg}} + \mu \cdot C_{\mathrm{param}},
\end{equation}
which is also
\begin{equation}
C =  C_{\mathrm{pix}} + \mu \cdot C_{\mathrm{param}}.
\label{eqn:final}
\end{equation}
%



\section{Algorithm in practice}
\label{handbook}

\subsection{Minimization}
\label{alg:minimi}

\begin{figure*}[ph!]
\centering
\includegraphics[width=17cm]{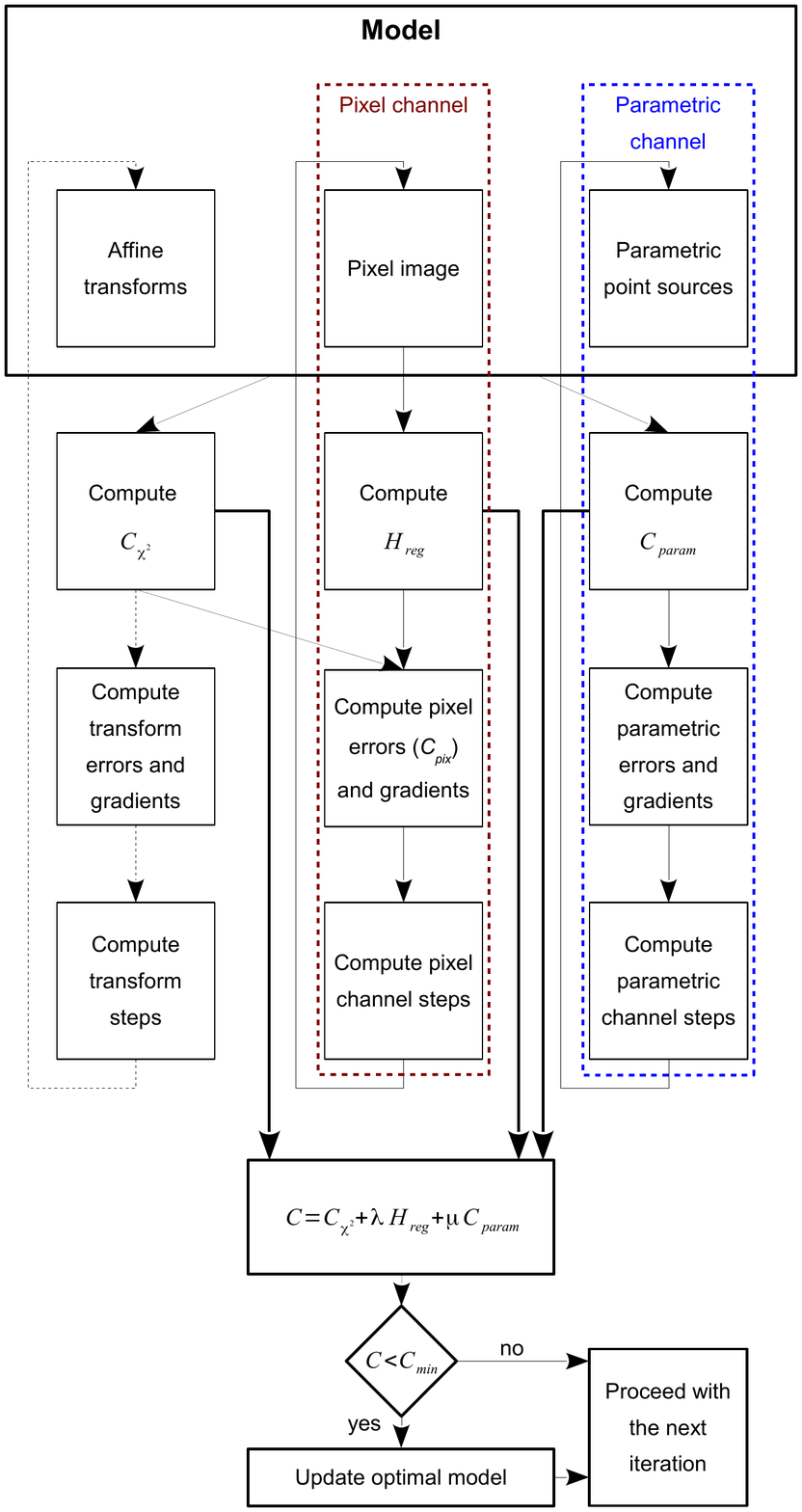}
\caption{Overview of one optimization step (``optimization'' box in Fig.\ref{fig:dec_flow}). The loop arrow for the affine transformation column is shown as a dotted line because the associated parameter update is not done at every iteration, but every few iterations. A typically number is one update every 30 iterations.}
\label{fig:minimi_flow}
\end{figure*}

The algorithm with which a cost function in high dimensions can
be minimized is a key element in a non-linear optimization. In our case, the difficulties arising from parameter degeneracies combine with a computationally intensive cost function motivated
by the use of a simple steepest-descent algorithm. However, we used a local estimation of the error for each parameter to compute the gradient, so that we did not need to compute the cost function on each parameter during the gradient estimation. This is possible because the influence of a small variation on a parameter will be visible mainly on small scales. The update steps for the parameters were computed from the corresponding terms of the cost function. This applies to each pixel $i$ of the pixel channel, but also to the point-source parameters: we cropped the residual image around the source at some adjustable radius before computing $C_{\mathrm{param}}$. However, the affine transform parameter steps are computed using the whole images. In Fig.~\ref{fig:minimi_flow} we show the different stages of one minimization step, illustrating the behavior of each channel and their effect on each other.

The algorithm was implemented in \texttt{Python}, with the possibi\-lity to use CUDA for the FFT computations, which uses graphi\-cal processing units (GPU). Without CUDA enabled, a typical deconvolution takes 15 seconds on a single processor for 1000 iterations for a $32\times 32$ image deconvolved to a final sampling of $64\times 64$ pixels. This computation time scales as $\mathcal{O}(\mathrm{n}^{2})$ with the image size and with the sampling factor. A more demanding deconvolution of a $64\times 64$ image with a sampling factor of 4 ($256\times 256$ deconvolved image) and 3000 iterations takes up to 10 minutes. However, most images require 100-300 iterations, and the use of CUDA typically improves the running time by a factor of 2 to 20, depending on the image size.

\subsection{Point-source detection}

An important element of our deconvolution algorithm is the initial identification of point sources and the estimation of sufficiently good initial conditions (position, intensity) to start the minimization process. Depending on the applications, it might be crucial to correctly distinguish between objects that are perfect point sources and those that are sharp, but not exactly point sources (e.g., compact galaxies). 

Evaluating the sharpness of an object can be achieved by measuring its n$^{th}$ derivatives and by comparing to the n$^{th}$ derivatives of the observed PSF. In practice, we took the Laplacian of the data and of the PSF and applied pattern matching using the fast normalized cross correlation proposed by \cite{Lewis1995}. This cross correlation is very efficient for a single-pass pattern matching, and the normalization it uses makes it independent of the source intensities. After the pattern matching, a simple thresholding operation gives the positions of the sources meeting the shape requirements. This process is summarized in Alg.~\ref{alg:findsources}, where $\star$ is the cross-correlation operator and \emph{MEANFILT} is a mean filter of the same size as the windowed PSF. 

\begin{algorithm}[!h]
\KwData{\emph{PSF}: windowed PSF; $D$: image; $n$: derivative order; t: threshold}
\KwResult{$l$: list of point-source positions}
\Begin{
        $l \longleftarrow []$\;
        $\mathit{PSF}' \longleftarrow \frac{\partial^n \mathit{PSF}}{\partial x^n}$\;
        $D' \longleftarrow \frac{\partial^n D}{\partial x^n}$\;
        $\mathit{corr} \longleftarrow \frac{D'\star \mathit{PSF}'}{\sqrt{\mathit{MEANFILT}(D'^2)}} $\;
        $\mathit{regions} \longleftarrow \mathit{GetRegions}(\mathit{corr} \geq t)$\;
        \For{$r \in regions$}{
                $p \longleftarrow \mathit{CenterOfMass}(r)$\;
                $l.\mathit{append}(p)$\;
        }
\KwRet{$l$}
}
\caption{Finding point sources}
\label{alg:findsources}
\end{algorithm}

Several strategies may be used to determine a good threshold value. We did this by iteratively raising the threshold on images simulated with the same noise characteristics as the original image until no false positive was detected. We adopted this threshold for the original image.

\begin{figure}[!h]
\centering
\includegraphics[width=\columnwidth]{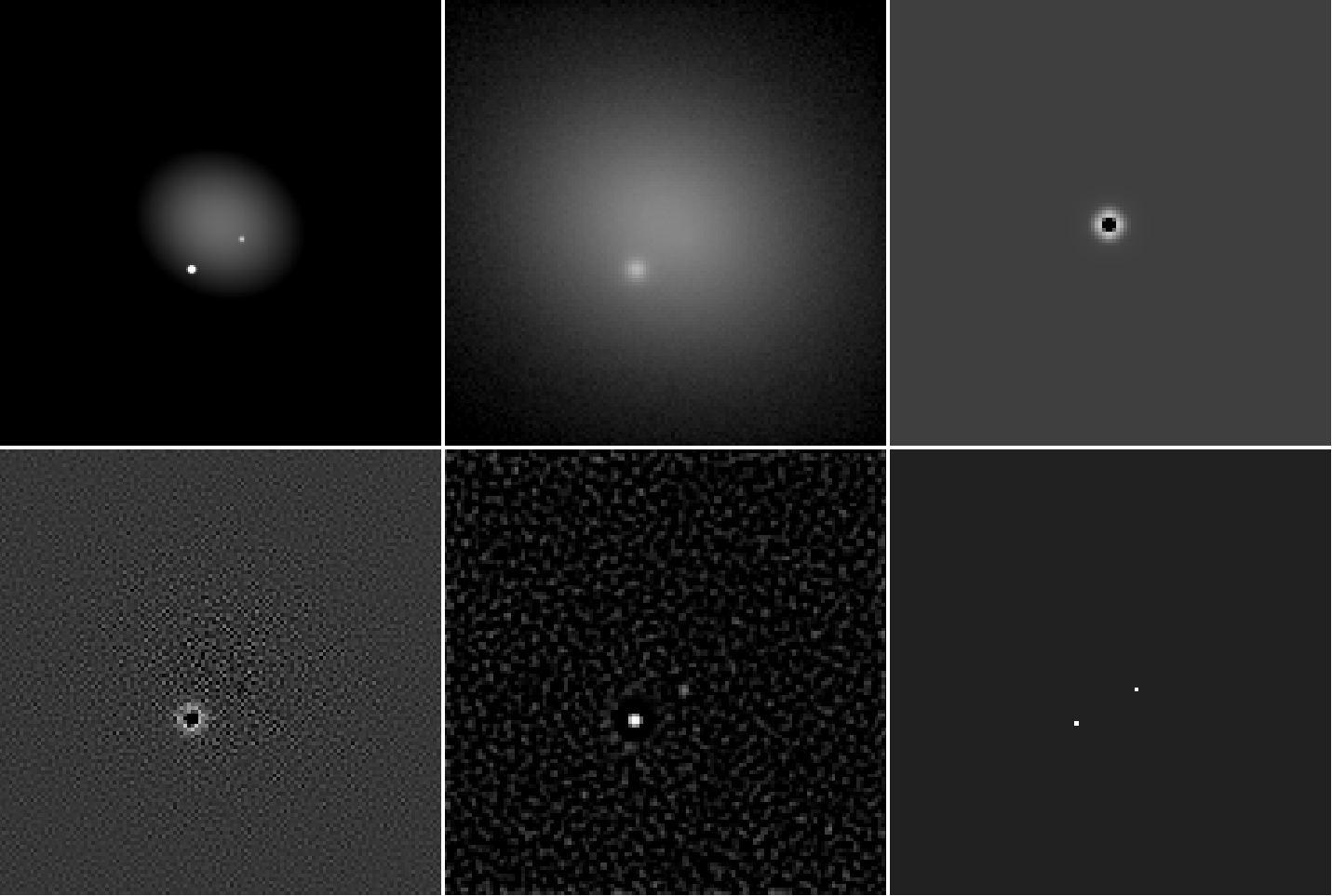}
\caption[]{Illustration of the different steps for detecting point sources. {\it Top:}  a truth image containing an extended source and two point sources is shown on the left. The middle panel shows the same image convolved with a PSF and affected by Poisson noise. The right panel shows the Laplacian of the PSF. {\it Bottom:} the Laplacian of the simulated PSF-convolved and noisy image is shown in the left. The middle panel shows its cross correlation by the Laplacian of the PSF. A thresholded version of this image is shown in the right panel, with the two point sources well detected. 
}
\label{fig:src_detect}
\end{figure}

We illustrate this process in Fig.~\ref{fig:src_detect}, using a simulated image containing two point sources superposed on an extended source. In this simulation one of the point sources is bright, the other one is faint and hidden in the glare of the extended source. Both point sources are identified by our algorithm.

\subsection{PSF construction}

The kernel $\mathbf{P}$ used in the deconvolution process has to be constructed for each data frame. Following the ideas of \citet{MCS} and Eq.~\ref{eqn:psf_mcs}, this kernel was obtained by deconvolving the observed PSF (that is, field stars) by the target PSF $\mathbf{g}$.
In practice, we did this in two successive steps. First, we fit an elliptical Moffat-like profile that gives the main characteristics of the PSF: width, ellipticity, orientation, and $\beta$ exponent (PSF "wings"). The fit was done simultaneously for several stars. Instead of fitting an exact Moffat profile, we approximated this function by a sum of three Gaussians. The ratios between the widths of these Gaussian profiles depend on $\beta$ and were empirically calibrated.
This makes it possible to analytically deconvole the PSF, hence greatly reducing the computational cost.

Second, the residuals from this analytical fit to the selected stars were also deconvolved by the target PSF $\mathbf{g}$. The affine transform introduced in Eq.~\ref{eq:afftrans} can be modified accordingly by setting $dx_j$ and $dy_j$ to the Moffat position inside the image stamp and $t_j$ to its intensity. In this manner, we scaled the reference residuals in such a way that they corresponded to a Moffat with a peak intensity $I_0 = 1$. The implemented algorithm allows the parameters of the affine transformations to vary during the minimization process, if necessary. The result of the process is a numerical array containing a lookup table correction to apply (by adding it) to the triple-Gaussian profile computed in the first step of the PSF construction. 

In these two steps, the algorithm includes the possibility to mask cosmic rays, CCD traps, close companions to the stars, or any undesired regions of the data frames. 




\begin{figure}[t!]
\centering
\includegraphics[width=6.5cm]{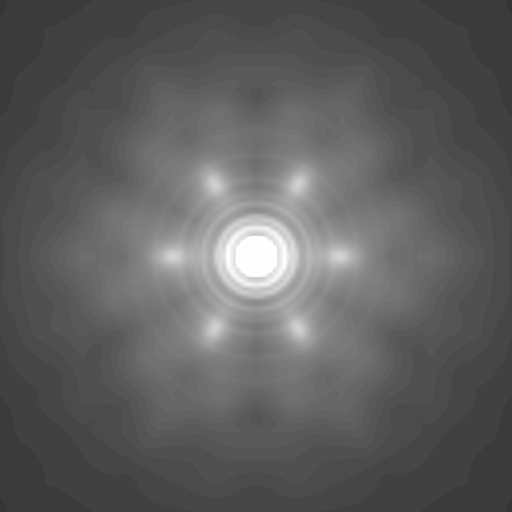}
\caption[E-ELT PSF]{PSF used for the photometric tests and the image simulation shown in Fig.~\ref{fig:eelt_dec}. We took here a particular complex PSF, for the MICADO instrument, which is
representative of the next generation of giant telescopes. The PSFs produced by space telescopes such as the HST or the JWST are qualitatively similar to that of the E-ELT.}
\label{fig:eelt_psf}
\end{figure}

\begin{figure*}[t!]
\centering
\includegraphics[width=6.1cm]{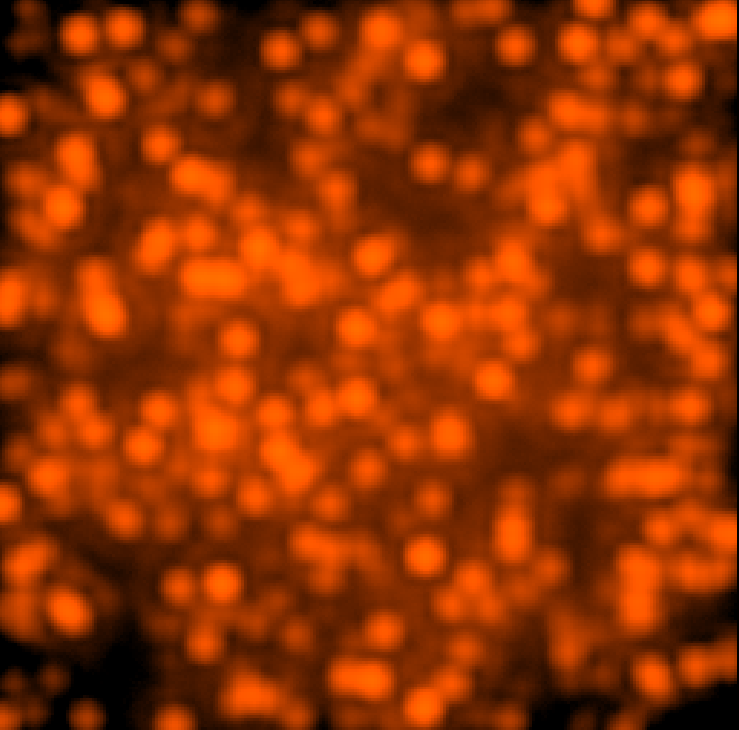}
\includegraphics[width=6cm]{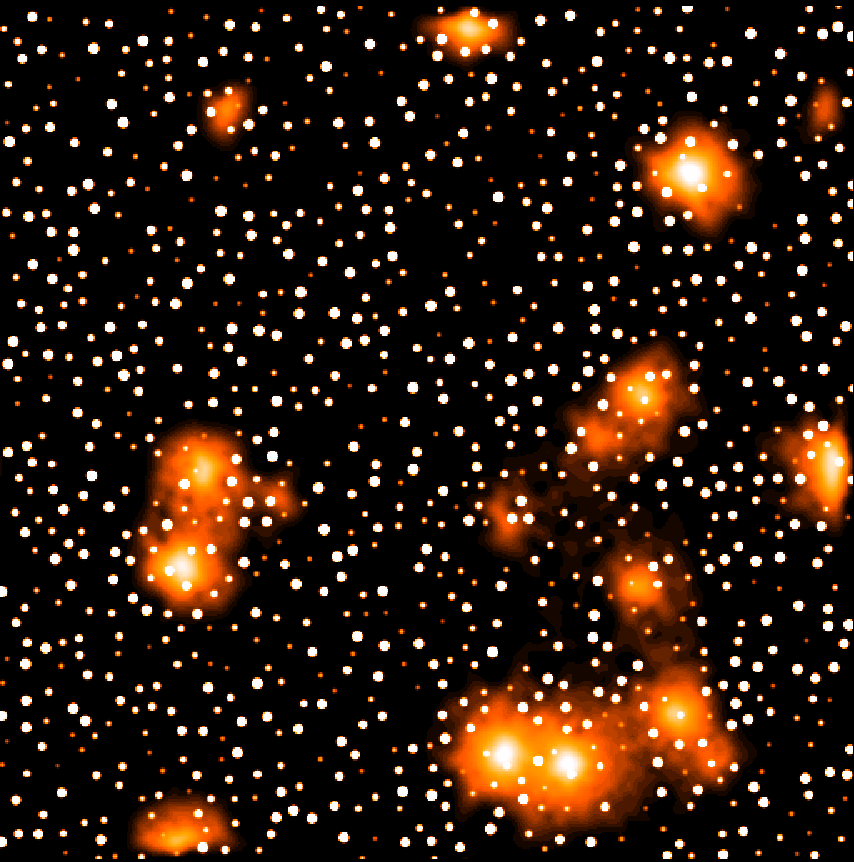}
\includegraphics[width=6cm]{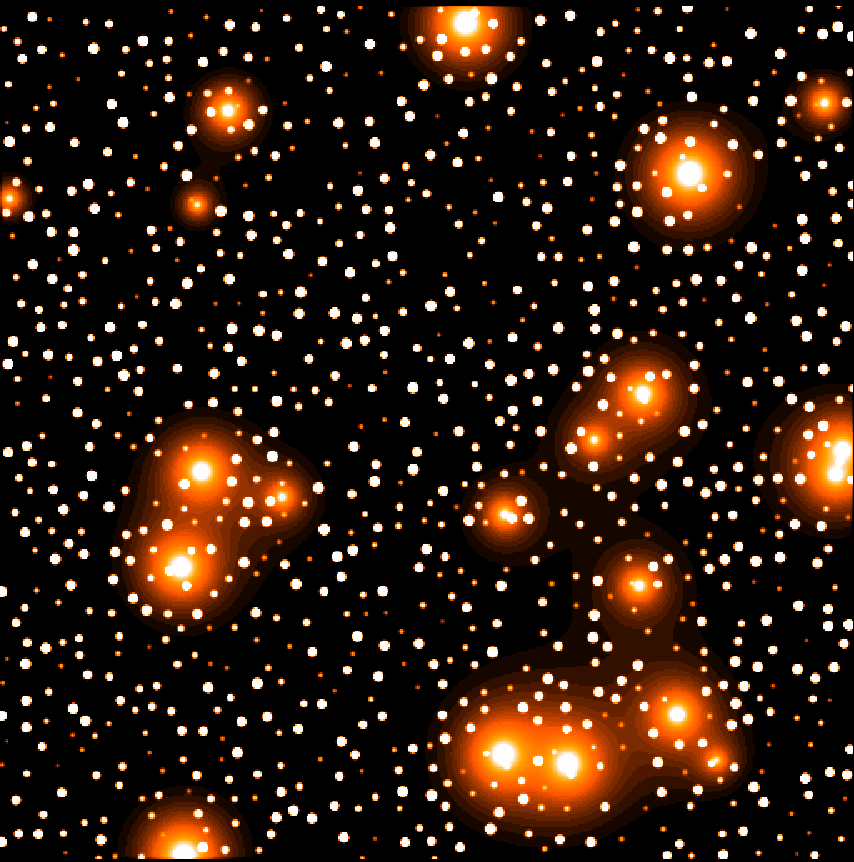}
\caption[Deconvolution example of an E-ELT crowded field image]{Example of deconvolution for a crowded field containing a mix of point
sources and extended sources. {\it Left:} simulated E-ELT image using the PSF available from ESO (Fig.~\ref{fig:eelt_psf}). The image contains 1000 randomly placed stars and 20 extended sources represented by elliptical Sersic profiles. {\it Middle:} the deconvolved image, where the PSF (i.e., the target PSF $\mathbf{g}$) is now a Gaussian with a FWHM of two pixels. The new linear pixel scale is twice smaller than the original. {\it Right:} truth image used to produce the simulated data in the left panel. The photometry of the point sources in this image is displayed in Fig.~\ref{fig:mag_dif}. The extended sources are recovered very
well, although they are completely hidden by the crowded field of PSFs in the original data.}\label{fig:eelt_dec}
\end{figure*}

\section{Tests with simulated and real data}
\label{tests}

In the following, we evaluate the performances of the algorithm in terms of photometric and astrometric accuracy with simulated images. We also study the ability of the algorithm to recover the shape of faint extended sources superposed with point sources, and we test how the method can help resolve and measure narrow blends of point sources under various contrast and separation conditions.

\subsection{Photometry and astrometry of point sources}

We tested the precision and accuracy of our algorithm for a crowded field of point sources that also contains extended sources. To do so, we simulated an image as may be obtained with the future E-ELT. The PSF used, computed by ESO for the MICADO instrument\footnote{\url{http://www.bo.astro.it/maory/Maory/PSF_06_2012.html}}, is shown in Fig.~\ref{fig:eelt_psf}. It was used to convolve a ``truth image'' containing 1000 point sources spanning a range of 11 magnitudes, and 20 extended sources spanning a range of 2 magnitudes. The typical width of the extended sources is $\mathrm{FWHM} \sim 10$ pixels. The simulated image size is 256 $\times$ 256 pixels. 

The deconvolution of the E-ELT image uses a linear pixel scale that is twice smaller than in the simulation, meaning that the deconvolved image is 512 pixels on a side. We considered the PSF to be exactly known, that is, we did not measure it from simulated stellar images. Instead, we used the full noise-free E-ELT PSF of Fig.~\ref{fig:eelt_psf} to construct the kernel $\mathbf{P}$ needed to reach a final resolution with a FWHM of two pixels (i.e., one pixel of the original data).

In our test, all magnitude scales are arbitrary, therefore we present our results as a function of $S/N$. The latter was computed as the total source flux over the total photon shot noise inside an aperture, as follows
\begin{equation}
S/N = F_{\mathrm{Tot}} \dfrac{1}{\sqrt{ \sum_{i=c-r}^{c+r} D_i}},
\label{eqn:snr}
\end{equation}
where $D_i$ is the pixel $i$ in the data, $c$ is the center of the aperture, and $r$ is its radius. For Gaussian sources with a FWHM of two pixels ($w=2$) and with a peak intensity, $a$, the total flux is
\begin{equation}
F_{\mathrm{Tot}} = 2\pi a \dfrac{w}{2\sqrt{2 \mathrm{ln}(2)}}
= a \dfrac{\pi}{\mathrm{ln}(2)}
.\end{equation}
\comment{malte}{Do we need this equation ? Everything works directly with fluxes, this $a$ stuff is hidden deep inside the code, no ? }
\comment{nico}{As you like. Fred?}\comment{Fred}{yes, keep it. It's obvious but sometimes useful}

We used an aperture with $r=20$ pixels for this computation, which corresponds to about twice the FWHM of the PSF of the original image. The deconvolved image  is shown in Fig.~\ref{fig:eelt_dec} along with the original image and the truth image.

\begin{figure}[t!]
\includegraphics[width=9.2cm]{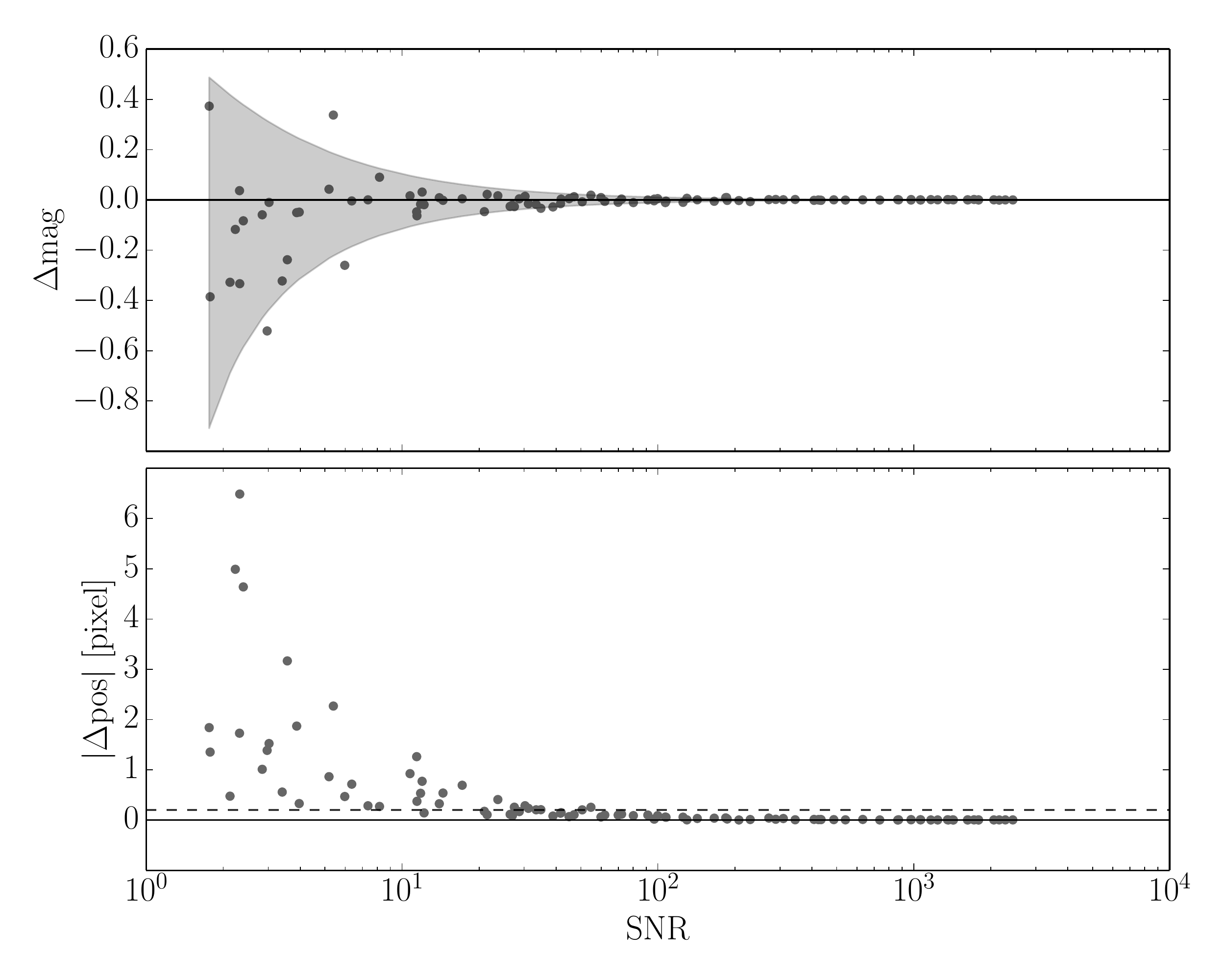}
\caption[Point-source photometry and astrometry using the E-ELT simulation]{Point-source photometry and astrometry using the E-ELT simulation shown in Fig.~\ref{fig:eelt_dec}. The differences in magnitude (top panel) and in position (bottom panel) with respect to the true values are shown as a function of S/N. The shaded area represents the photometric uncertainty in the data, as expected from the photon shot-noise. In the lower panel the astrometric accuracy is given in pixel units of the original image. \comment{malte}{Bloody MCS heritage ! Shouldn't we speak only in terms of pixels of the original data, or in arcsec ? Otherwise will have to specify which pixels all across this section...} The dotted line shows the 0.2 pixel limit. Most of the point sources with $S/N > 20$ are below this line. 
}
\label{fig:mag_dif}
\end{figure}

Our photometric and astrometric results are summarized in Fig.~\ref{fig:mag_dif}. This test, carried out on a realistic image of a fairly crowded field, shows that our photometry is limited by the photon-noise and is not biased. It also shows that most of the point sources with $S/N > 20$ have an astrometric uncertainty of about a tenth of the pixel size of the original data, which is remarkable given the crowding of the field.
\comment{malte}{seems to be much larger in the figure, check !}\comment{nico}{indeed... I don't know how this sentence came to be. Commented}\comment{malte}{ok, removed mention to small pixels and fixed sentence}



\begin{figure}[t!]
\includegraphics[width=9.2cm]{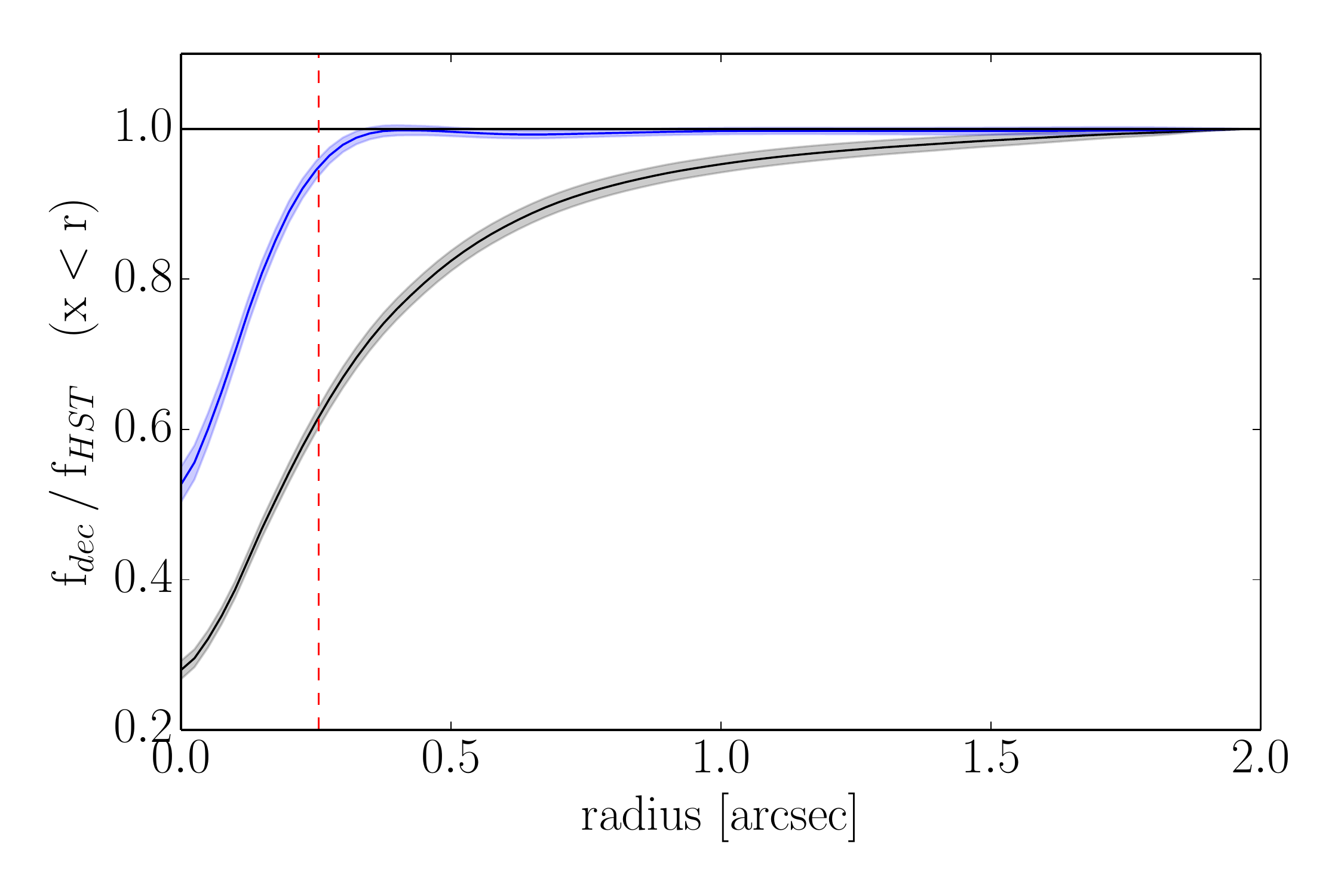}
\caption[Radial profile error]{Ratio between the radial light profile of EDisCS galaxies as measured on the deconvolved VLT data and as measured on the HST data (blue). The same ratio is shown when the radial profile is measured on the original VLT data (black). The filled regions around the lines are the $3\sigma$ deviations around the mean value for the 143 galaxies. The red dotted line indicates the 95\% flux limit fidelity for the deconvolved VLT data.}
\label{fig:aper_prof}
\end{figure}

\subsection{Shape of extended sources}

We now test the ability of our algorithm to deconvolve extended sources, that is, those
sources containing spatial frequencies lower than those contained in {$\mathbf
  g$}, the target PSF. This is relatively simple for objects containing
frequencies much lower than {$\mathbf g,$} although, as we showed in
Sect.~\ref{wavelet}, the performances of the algorithm depend on the choice of
the regularization term.

In practice, however, sharp objects with frequencies similar to but slightly
lower than {$\mathbf g$} must be handled with particular care. One obvious
solution is to analytically model such objects, introducing a third channel in the
image deconvolution algorithm. This is equivalent to a full forward fitting, as
implemented in {\tt galfit} \citep{Peng2011}, for example. Our goal is instead to avoid
introducing any prior on the shape of the objects so that any arbitrary shape can
be deconvolved. We show in the following that our algorithm model such
objects.

A typical astrophysical situation where compact objects are considered is in the
field of multi-band imaging of distant galaxies. The latter are often similar
in angular size to the PSF width, but they are more extended than the PSF.

In the following, we devise a test to evaluate how our deconvolution algorithm
performs aperture photometry for such small objects. To do so, we used a sample
of galaxies gathered from the ESO Distant Cluster Survey \citep{White2005}. A
total of 143 elliptical and S0 galaxies were selected, with spectroscopic
redshifts \citep{Halliday2004, Milvang-JensenB.2008} and visual morphologies based
on HST/ACS/F814W images \citep{Desai2007b}.  Our test consists of deconvolving
their VLT/FORS2 $I$-band images down to the HST/ACS resolution and to compare the
photometric measurements carried out on these images and on the HST images.

The typical seeing of the ground-based data is 0.5-0.8\arcsec, which we deconvolved
down to a spatial resolution of 0.1\arcsec, which is approximately the resolution of the HST in
this band. We then derived the galaxy radial light profile by successive photometric
measurements in circular apertures of increasing radius and compared this radial
profile to the one measured on the HST image. The result of this procedure is shown
in Fig.~\ref{fig:aper_prof}, where we display the mean and the 3$\sigma$ deviation of
the ratios of the two profiles for the full set of our galaxy sample. This indicates
whether or not the profiles deviate from each other, where they
do this, and by how much.  For
comparison and to estimate the gain of the deconvolution procedure, we also compared
the HST profiles to those measured on the original ground-based images.  It is clear
that the original VLT/FORS2 images do not allow measuring the profiles of
the galaxies correctly. The two sets of profiles deviate until 1.5\arcsec\ away from the galaxy
centers, which is very close to the size of the galaxies at this redshift. In
contrast, the deconvolved VLT data and the HST data agree to much better than 5\%
down to 0.4\arcsec\ from the galaxy center.  Below 0.25\arcsec, the profile deviates
dramatically. However, this region is very small and has approximately the size of a
FORS2 pixel (0.2\arcsec). Moreover, at such small distances from the galaxies' centers,
the HST PSF itself significantly affects our experiment.

While this test was carried out on early-type galaxies, we focus in a companion paper
\citep{Cantale2016} on deconvolving the late-type galaxies of
the EDisCS sample, with the goal of separating the bulge and disk components for
cluster and field galaxies at redshift $0.5 < z < 0.8$.

\subsection{Point-source deblending}

Another common situation where a deconvolution algorithm applies is to separate extremely blended point sources, especially if the contrast between the sources is large. Typical astrophysical situations where this occurs are the astrometric measurement of multiple stars, the search for planetary companions of bright stars, and the separation of the lensed images of distant quasars.

We tested this situation by generating pairs of point sources with random separations and contrasts. The PSF used for this experiment was a Moffat profile with a FWHM of 7.5 pixels that mimics a typical seeing of 0.75\arcsec\ for a ground-based telescope with 0.1\arcsec\ per pixel. We produced 1000 such pairs that we deconvolved with our algorithm. The target resolution was 0.2\arcsec\ and the pixel size on the deconvolved images was the same as on the original image.

\begin{figure*}[!t]
\centering
\includegraphics[width=14cm]{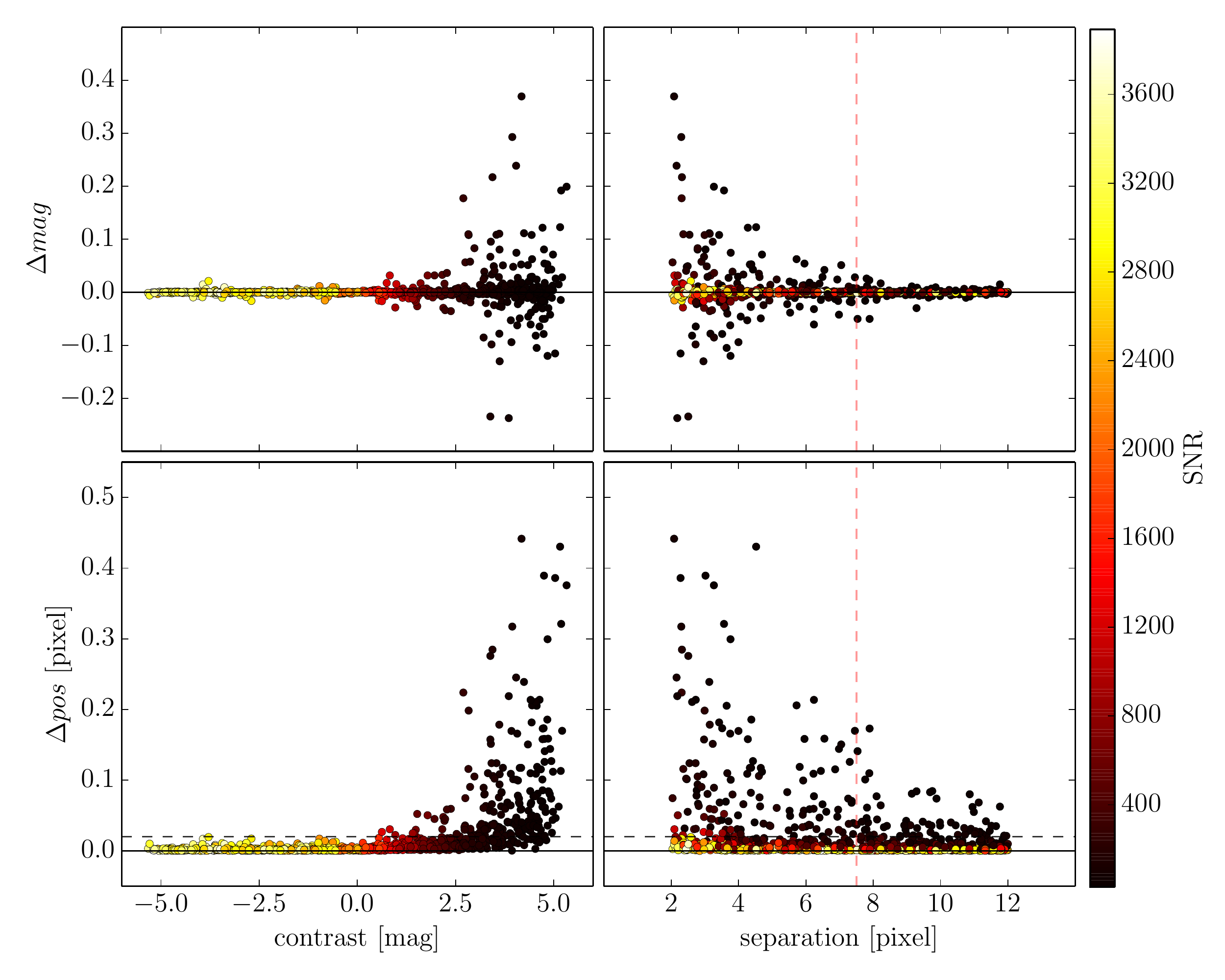}
 \caption{Illustration of the deblending capability of our algorithm for pairs of  point sources. The upper panels show the photometric accuracy as a function of contrast between the objects and as a function of separation. The astrometric precision is given in the lower panels in units of deconvolved pixels. The vertical red dashed line indicates the FWHM of the PSF in the simulation (7.5 pixels). The horizontal black dashed line indicates a 0.02 pixel precision for reference. The color code represents the S/N of the objects as defined in Eq.~\ref{eqn:snr}.}
\label{fig:deblend}
\end{figure*}

Figure~\ref{fig:deblend} displays the error on the recovered photometry and astrometry as a function of contrast and as a function of separation between the point sources in the simulation. 

The images were constructed by choosing the magnitude of the first source randomly in the range $-17 < m < -18$ and at a random subpixel position. A random separation (between 2 and 12 pixels) was computed for the second source, which was then placed, also at random, on a circle of that radius. The magnitude of the second (fainter) source was also chosen with a contrast of up to 6 magnitudes. The image was then convolved by the PSF and degraded by Poisson noise, mimicking a sky level of 1000 ADUs. Initial conditions were set by adding a random offset of 0.5 pixels to the true position and of 10\% to the original flux.
In this specific example, we assumed that the point source detection was provided by other means. With real data, the image with no point source at all in the parametric channel may well be deconvolved before detecting point-source candidates on this first deconvolved image, and then deconvolving the full two channels. This has been done in the past with the orginal MCS algorithm applied to HST images of globular clusters seen in projection on the bulge of M31 by \cite{jablonka2000}. If necessary, point sources may be added or removed to carry out yet another deconvolution
pass. Whether a point source must be added or removed can be decided by inspecting the residual map (image of the $\chi^2$)
and determining regions that are under- or overfitted. 

In Fig.~\ref{fig:deblend} each member of each pair is displayed, with the color code indicating the $S/N$. The separation and contrast are shown with respect to other member of the pair. The astrometry of sources with contrast as high as 5 magnitudes is recovered down to 0.02 pixels (dotted black line) as soon as the $S/N$ is higher than 150. For $S/N$ higher than 200, the astrometry and photometry are recovered down to 0.02 mag and 0.01 pixels, regardless of the contrast and separation. For low $S/N$, the photometric precision is limited by the photon noise from the bright member, and the photometry and subpixel astrometry of the low $S/N$ member become uncertain with contrasts higher than four magnitudes. However, science cases with these configurations would usually require the efficient characterization of the brightest source alone to study the companion after subtraction. Except for sources with the lowest S/N, the photometric and astrometric precision is good even at separations smaller than a PSF with a FWHM size of 7.5 pixels.

\section{Conclusion}
\label{conclusion}

We present a new image deconvolution algorithm that was developed
from the foundations of the MCS image deconvolution algorithm \citep{MCS}: a sampled image should not be deconvolved using the observed PSF, but  with a narrower PSF, which ensures that the sampling theorem is always respected. 

Using this narrower PSF, the algorithm decomposes the data into two channels. The first contains the point sources in the image, the second contains all extended sources. All point sources in the original data are transformed into analytical circular Gaussians for which the photometry and the astrometry are given as outputs. In addition, the algorithm can simultaneously deconvolve many dithered frames of the same object. An output of the deconvolution process is therefore the light curve of any photometrically variable point source and a deep sharp combined image decomposed into two channels. In the present work we introduced significant modifications to the original algorithm:

\begin{itemize}

\item a new regularization term based on image de-noising using wavelet decomposition of the pixel channel of the deconvolved image and cycle-spinning. This results in an adaptive regularization that accounts for the variable S/N across the data and for data of a wider dynamical range than with the original MCS algorithm.  We note that this choice can be improved in many ways, for instance,
using sparse regularization, but our de-noising scheme delivers satisfactory results. Future developments may explore other regularization schemes.

\item a method for minimizing the degeneracies between the parametric and the pixel channels in the vicinity of point sources. This is done by minimizing the norm of the $n$-th order derivatives of the residual image. For low S/N data, wavelet de-noising can be applied to these derivative images if needed.

\item a general affine transformation that allows simultaneously
deconvolving images that are affected by distortions. This can also be critical in building the deconvolution kernel from stars located at very different places on the detector that are therefore distorted in different ways. 

\item a recipe to derive the noise per pixel (noise map) when the data are very noisy. This recipe is based on estimating the noise with the help of a de-noised version of the data instead
of on the data themselves. 

\end{itemize}

We then showed from simulated images that our algorithm can perform photon-noise limited photometry and  reliable astrometry in crowded fields and for very narrow blends of point sources with high contrast. This was tested even for complex diffraction limited PSFs, such as those of space telescopes (HST, JWST) or giant ground-based telescopes operating with AO (the E-ELT and TMT).

The MCS algorithm has been applied to a broad range of applications in the past
\cite[e.g.,][]{Eigenbrod2008, Gillon2006, Magain2005, Meylan2005, Letawe2004}. The
improved version presented here has been used to obtain accurate light curves
of the blended images of gravitationally lensed quasars monitored by the COSMOGRAIL
program \citep[e.g.,][]{Tewes2013, Rathna2013}. It was recently applied to disk galaxies at intermediate redshift, for which it sucessfully allowed the derivation
of colors, as reported by \citet{Cantale2016}.

Applications of the new regularization
are possible in the field of multiband photometry with matched seeing, which is vital to study galaxy formation and evolution. Current algorithms for carrying out multiband photometry usually involve model fitting \citep[e.g.,][]{Megamorph}. With our deconvolution method, the pixel channel of the resulting image can have any arbitrary shape and allow more flexibility than model fitting.
Other applications are possible in the field of supernovae light curves, where the supernova must be measured free of any contamination by the host galaxy and free of any PSF effect. Image deconvolution will become of crucial importance with future giant telescopes operating exclusively with AO and displaying complex diffraction-limited PSFs. This type of observations will require further development of our algorithm. In particular, the PSF will need to be modified iteratively during the deconvolution process (myopic deconvolution) to account for the chromatic dependence of the PSF and for its spatial variability across the field of view. For standard non-AO ground-based telescopes, our algorithm can produce images that
are similar in resolution to the HST images. This requires that
the PSF is stable across the field, a fairly good seeing of 0.6-0.8\arcsec, and a pixel size between 0.1\arcsec\ and 0.2\arcsec, as produced with the current imagers mounted on typical 10m telescopes. Such data may well complement single-band HST data.

\begin{acknowledgements}
This research is supported by the Swiss National Science Foundation (SNSF). MT acknowledges support from a fellowship of the Alexander von Humboldt Foundation and from the DFG grant Hi 1495/2-1.
\end{acknowledgements}

\bibliographystyle{aa}
\bibliography{bib}

\begin{thebibliography}{43}
\expandafter\ifx\csname natexlab\endcsname\relax\def\natexlab#1{#1}\fi

\bibitem[{Aumont \& Macias-Perez(2007)}]{Aumont2007}
Aumont, J. \& Macias-Perez, J.~F. 2007, \mnras, 376, 739

\bibitem[{Baccigalupi {et~al.}(2000)Baccigalupi, Bedini, Burigana, {De Zotti},
  Farusi, Maino, Maris, Perrotta, Salerno, Toffolatti, \&
  Tonazzini}]{Baccigalupi2000}
Baccigalupi, C., Bedini, L., Burigana, C., {et~al.} 2000, \mnras, 318, 769

\bibitem[{Bobin {et~al.}(2013)Bobin, Starck, Sureau, \& Basak}]{Bobin2013}
Bobin, J., Starck, J.-L., Sureau, F., \& Basak, S. 2013, \aap, 550, A73

\bibitem[{Bontekoe {et~al.}(1993)Bontekoe, Koper, \& Kester}]{Bontekoe1993}
Bontekoe, T., Koper, K., \& Kester, D. 1993, \aap, 284, 1037

\bibitem[{{Cantale} {et~al.}(2016){Cantale}, {Jablonka}, {Courbin}, {Rudnick},
  {Zaritsky}, {Meylan}, {Desai}, {De Lucia}, {Aragon-Salamanca}, {Poggianti},
  {Finn}, \& {Simard}}]{Cantale2016}
{Cantale}, N., {Jablonka}, P., {Courbin}, F., {et~al.} 2016, ArXiv1601.05192

\bibitem[{Coifman \& Donoho(1995)}]{Coifman1995}
Coifman, R.~R. \& Donoho, D.~L. 1995, Time, 103, 125

\bibitem[{Desai {et~al.}(2007)Desai, Dalcanton, Aragon-Salamanca, Jablonka,
  Poggianti, Gogarten, Simard, Milvang-Jensen, Rudnick, Zaritsky, Clowe,
  Halliday, Pello, Saglia, \& White}]{Desai2007b}
Desai, V., Dalcanton, J.~J., Aragon-Salamanca, A., {et~al.} 2007, ApJ, 660,
  1151

\bibitem[{{Eigenbrod} {et~al.}(2008){Eigenbrod}, {Courbin}, {Sluse}, {Meylan},
  \& {Agol}}]{Eigenbrod2008}
{Eigenbrod}, A., {Courbin}, F., {Sluse}, D., {Meylan}, G., \& {Agol}, E. 2008,
  \aap, 480, 647

\bibitem[{{Fruchter} \& {Hook}(2002)}]{drizzle}
{Fruchter}, A.~S. \& {Hook}, R.~N. 2002, \pasp, 114, 144

\bibitem[{{Gillon} {et~al.}(2006){Gillon}, {Pont}, {Moutou}, {Bouchy},
  {Courbin}, {Sohy}, \& {Magain}}]{Gillon2006}
{Gillon}, M., {Pont}, F., {Moutou}, C., {et~al.} 2006, \aap, 459, 249

\bibitem[{{Giovannelli} \& {Coulais}(2005)}]{Giovannelli2005}
{Giovannelli}, J.-F. \& {Coulais}, A. 2005, \aap, 439, 401

\bibitem[{Halliday {et~al.}(2004)Halliday, Milvang-Jensen, Poirier, Poggianti,
  Jablonka, Aragon-Salamanca, Saglia, {De Lucia}, Pello, Simard, Clowe,
  Rudnick, Dalcanton, White, Zaritsky, Aragon-Salamanca, \&
  Pello}]{Halliday2004}
Halliday, C., Milvang-Jensen, B., Poirier, S., {et~al.} 2004, A\&A, 427, 397

\bibitem[{{H{\"a}u{\ss}ler} {et~al.}(2013){H{\"a}u{\ss}ler}, {Bamford}, {Vika},
  {Rojas}, {Barden}, {Kelvin}, {Alpaslan}, {Robotham}, {Driver}, {Baldry},
  {Brough}, {Hopkins}, {Liske}, {Nichol}, {Popescu}, \& {Tuffs}}]{Megamorph}
{H{\"a}u{\ss}ler}, B., {Bamford}, S.~P., {Vika}, M., {et~al.} 2013, \mnras,
  430, 330

\bibitem[{Herranz {et~al.}(2009)Herranz, L\'{o}pez-Caniego, Sanz, \&
  Gonz\'{a}lez-Nuevo}]{Herranz2009}
Herranz, D., L\'{o}pez-Caniego, M., Sanz, J.~L., \& Gonz\'{a}lez-Nuevo, J.
  2009, \mnras, 394, 510

\bibitem[{H\"{o}gbom(1974)}]{Hogbom1974}
H\"{o}gbom, J. 1974, \aaps

\bibitem[{Hook \& Lucy(1994)}]{Hook1994a}
Hook, R. \& Lucy, L. 1994, in {The Restoration of HST Images and Spectra-II},
  Vol.~1, 86

\bibitem[{Hurier {et~al.}(2013)Hurier, Mac\'{\i}as-P\'{e}rez, \&
  Hildebrandt}]{Hurier2013}
Hurier, G., Mac\'{\i}as-P\'{e}rez, J.~F., \& Hildebrandt, S. 2013, \aap, 558,
  A118

\bibitem[{{Jablonka} {et~al.}(2000){Jablonka}, {Courbin}, {Meylan},
  {Sarajedini}, {Bridges}, \& {Magain}}]{jablonka2000}
{Jablonka}, P., {Courbin}, F., {Meylan}, G., {et~al.} 2000, \aap, 359, 131

\bibitem[{Kacprzak {et~al.}(2013)Kacprzak, Zuntz, Rowe, Bridle, Refregier,
  Amara, Voigt, \& Hirsch}]{Kacprzak2012}
Kacprzak, T., Zuntz, J., Rowe, B., {et~al.} 2013, \mnras, 427, 2711

\bibitem[{{Lefkimmiatis} \& {Unser}(2013)}]{Lefkimmiatis2013}
{Lefkimmiatis}, S. \& {Unser}, M. 2013, IEEE Transactions on Image Processing,
  22, 4314

\bibitem[{{Letawe} {et~al.}(2004){Letawe}, {Courbin}, {Magain}, {Hilker},
  {Jablonka}, {Jahnke}, \& {Wisotzki}}]{Letawe2004}
{Letawe}, G., {Courbin}, F., {Magain}, P., {et~al.} 2004, \aap, 424, 455

\bibitem[{Lewis(1995)}]{Lewis1995}
Lewis, J. 1995, Vision interface, 1995

\bibitem[{Lucy(1974)}]{Lucy1974}
Lucy, L.~B. 1974, \aj, 79, 745

\bibitem[{Luisier {et~al.}(2011)Luisier, Blu, \& Unser}]{Luisier2011}
Luisier, F., Blu, T., \& Unser, M. 2011, IEEE transactions on image processing
  : a publication of the IEEE Signal Processing Society, 20, 696

\bibitem[{Magain {et~al.}(1998)Magain, Courbin, \& Sohy}]{MCS}
Magain, P., Courbin, F., \& Sohy, S. 1998, \apj, 494, 472

\bibitem[{{Magain} {et~al.}(2005){Magain}, {Letawe}, {Courbin}, {Jablonka},
  {Jahnke}, {Meylan}, \& {Wisotzki}}]{Magain2005}
{Magain}, P., {Letawe}, G., {Courbin}, F., {et~al.} 2005, \nat, 437, 381

\bibitem[{{Meylan} {et~al.}(2005){Meylan}, {Courbin}, {Lidman}, {Kneib}, \&
  {Tacconi-Garman}}]{Meylan2005}
{Meylan}, G., {Courbin}, F., {Lidman}, C., {Kneib}, J.-P., \& {Tacconi-Garman},
  L.~E. 2005, \aap, 438, L37

\bibitem[{Milvang-Jensen {et~al.}(2008)Milvang-Jensen, Noll, Halliday,
  Poggianti, Jablonka, Arag\'{o}n-Salamanca, Saglia, Nowak, von~der Linden, {De
  Lucia}, Pell\'{o}, Moustakas, Poirier, Bamford, Clowe, Dalcanton, Rudnick,
  Simard, White, \& Zaritsky}]{Milvang-JensenB.2008}
Milvang-Jensen, B., Noll, S., Halliday, C., {et~al.} 2008, A\&A, 482, 419

\bibitem[{{Peng} {et~al.}(2011){Peng}, {Ho}, {Impey}, \& {Rix}}]{Peng2011}
{Peng}, C.~Y., {Ho}, L.~C., {Impey}, C.~D., \& {Rix}, H.-W. 2011, {GALFIT:
  Detailed Structural Decomposition of Galaxy Images}, ascl1104.010

\bibitem[{Pirzkal {et~al.}(2000)Pirzkal, Hook, \& Lucy}]{Pirzkal2000}
Pirzkal, N., Hook, R., \& Lucy, L. 2000, in {Astronomical Data Analysis
  Software and Systems IX}, Vol. 216, 655

\bibitem[{{Rathna Kumar} {et~al.}(2013){Rathna Kumar}, {Tewes}, {Stalin},
  {Courbin}, {Asfandiyarov}, {Meylan}, {Eulaers}, {Prabhu}, {Magain}, {Van
  Winckel}, \& {Ehgamberdiev}}]{Rathna2013}
{Rathna Kumar}, S., {Tewes}, M., {Stalin}, C.~S., {et~al.} 2013, \aap, 557, A44

\bibitem[{Refregier {et~al.}(2012)Refregier, Kacprzak, Amara, Bridle, \&
  Rowe}]{Refregier2012}
Refregier, A., Kacprzak, T., Amara, A., Bridle, S., \& Rowe, B. 2012, \mnras,
  425, 1951

\bibitem[{Richardson(1972)}]{Richardson1972}
Richardson, W.~H. 1972, Journal of the Optical Society of America, 62, 55

\bibitem[{Rudin {et~al.}(1992)Rudin, Osher, \& Fatemi}]{Rudin1992}
Rudin, L.~I., Osher, S., \& Fatemi, E. 1992, Physica D, 60, 259

\bibitem[{{Selig} \& {En{\ss}lin}(2015)}]{Selig2015}
{Selig}, M. \& {En{\ss}lin}, T.~A. 2015, \aap, 574, A74

\bibitem[{Skilling \& Bryan(1984)}]{Skilling1984}
Skilling, J. \& Bryan, R. 1984, \mnras, 211

\bibitem[{Starck \& Murtagh(2007)}]{starckbook2007}
Starck, J.-L. \& Murtagh, F. 2007, Astronomical image and data analysis
  (Springer Science \& Business Media)

\bibitem[{Starck {et~al.}(2010)Starck, Murtagh, \& Fadili}]{starckbook2010}
Starck, J.-L., Murtagh, F., \& Fadili, M. 2010, Sparse Image and Signal
  Processing (Cambridge University Press)

\bibitem[{{Tewes} {et~al.}(2013){Tewes}, {Courbin}, {Meylan}, {Kochanek},
  {Eulaers}, {Cantale}, {Mosquera}, {Magain}, {Van Winckel}, {Sluse},
  {Cataldi}, {V{\"o}r{\"o}s}, \& {Dye}}]{Tewes2013}
{Tewes}, M., {Courbin}, F., {Meylan}, G., {et~al.} 2013, \aap, 556, A22

\bibitem[{Velusamy {et~al.}(2008)Velusamy, Marsh, Beichman, Backus, \&
  Thompson}]{Velusamy2008}
Velusamy, T., Marsh, K.~a., Beichman, C.~a., Backus, C.~R., \& Thompson, T.~J.
  2008, The Astronomical Journal, 136, 197

\bibitem[{Weir(1992)}]{Weir1992}
Weir, N. 1992, Astronomical Data Analysis Software and Systems I

\bibitem[{{White} {et~al.}(2005){White}, {Clowe}, {Simard}, {Rudnick}, {De
  Lucia}, {Arag{\'o}n-Salamanca}, {Bender}, {Best}, {Bremer}, {Charlot},
  {Dalcanton}, {Dantel}, {Desai}, {Fort}, {Halliday}, {Jablonka}, {Kauffmann},
  {Mellier}, {Milvang-Jensen}, {Pell{\'o}}, {Poggianti}, {Poirier},
  {Rottgering}, {Saglia}, {Schneider}, \& {Zaritsky}}]{White2005}
{White}, S.~D.~M., {Clowe}, D.~I., {Simard}, L., {et~al.} 2005, \aap, 444, 365

\bibitem[{Zhang {et~al.}(2008)Zhang, Fadili, Starck, \& Digel}]{Zhang}
Zhang, B., Fadili, M.~J., Starck, J.-L., \& Digel, S. 2008, Statistical
  Methodology, 5, 387

\end{thebibliography}

\begin{appendix}

\section{Note on the noise estimation}
\label{noise}

Our algorithm
requires an estimate of the noise amplitude in every pixel of the data. Any spatially homogeneous noise level is straightforward to estimate by analyzing empty areas of the data frames.
Astronomical images are, however, mainly affected by Poisson noise, which is usually approximated by a Gaussian noise with a standard deviation that varies with the square root of the signal. This is  true for our algorithm as well, since our regularization term is built on the fact that the noise is proportional to the square root of the signal and since astronomical images usually display a wide dynamical range. 

Estimating the amplitude of the noise per pixel, $\sigma_i$, in an image is challenging. This is usually done from the data themselves, $D_i$, following 
\begin{equation}
\sigma_i = \sqrt{D_i + \sigma_{sky}^2},
\label{shotnoise}
\end{equation}
where $\sigma_{sky}$ is the rms noise in the sky background. This calculation is valid for a frame where the sky background has been subtracted.

\begin{figure}[t!]
\includegraphics[width=8.8cm]{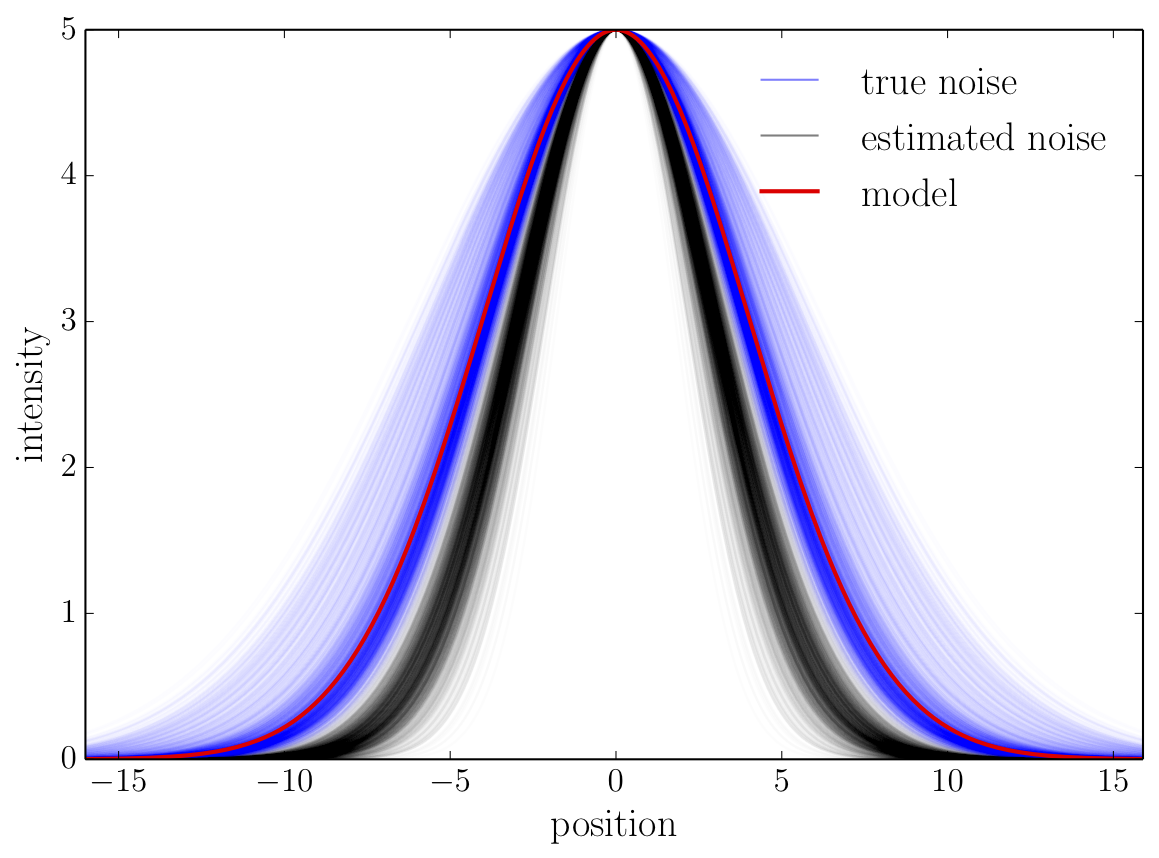}
\caption[]{Effect of an incorrect estimate of the Poisson noise on a very simple fit. A symmetrical Gaussian profile is simulated (red) and fitted for 1000 realizations of the simulation using the correct noise estimate (blue). When the error bars on the pixel values are computed using the (noisy) data themselves, the fit (black) is affected by systematics, meaning that in this case the FWHM of the Gaussian is underestimated.}
\label{fig:sigmabias}
\end{figure}

However, estimating the noise in this way, directly from the noisy data, results in a bias that can significantly affect any subsequent measurement. This bias is similar to the so-called noise-bias seen in weak-lensing measurements \citep[e.g.,][]{Refregier2012, Kacprzak2012}, but is different in nature and in amplitude. It can be understood by carrying out a very simple experiment in
which a constant flux level is attempted to be
fit to data points that are distributed around a constant value
and in which the data points are affected by Poisson noise. If the rms noise is computed from such noisy data using Eq.~\ref{shotnoise}, all points brighter than the mean level have overestimated error bars and all points fainter than the mean have underestimated error bars. As a result, the fit of a constant (flat line) to the data is biased downward with respect to the true zero level. Figure~\ref{fig:sigmabias} illustrates this in the slightly more complex case of a symmetrical Gaussian. In the figure, the fit is carried out in two ways: 1) using the exact noise estimate, that is, using the noise-free data to feed Eq.~\ref{shotnoise}, and 2) using the noise estimate carried out on the noisy data. The typical S/N for one realization of the data is about $S/N\sim 15$. In this experiment only the width of the Gaussian is a free parameter. It is biased downward when the shot noise is estimated from the noisy data.

To quantify the effect further, we carried out the same experiments as in \citet{Refregier2012}, but now with Poisson noise. This experiment consists of fitting a two-dimensional circular Gaussian profile with known intensity on a simulated two-dimensional circular Gaussian. Only the width of the Gaussian is fitted, all other parameters are exactly known. The fit was made twice, with and without a convolution kernel (PSF). We used the exact same simulation as in \cite{Refregier2012}: the width of the simulated Gaussian was $a_{\rm gal}=4$ pixels and the width of the PSF, when applicable, was $a_{\rm PSF}=5.33$ pixels. We also used the same definition of the S/N. 

\begin{figure}[t!]
\includegraphics[width=9cm]{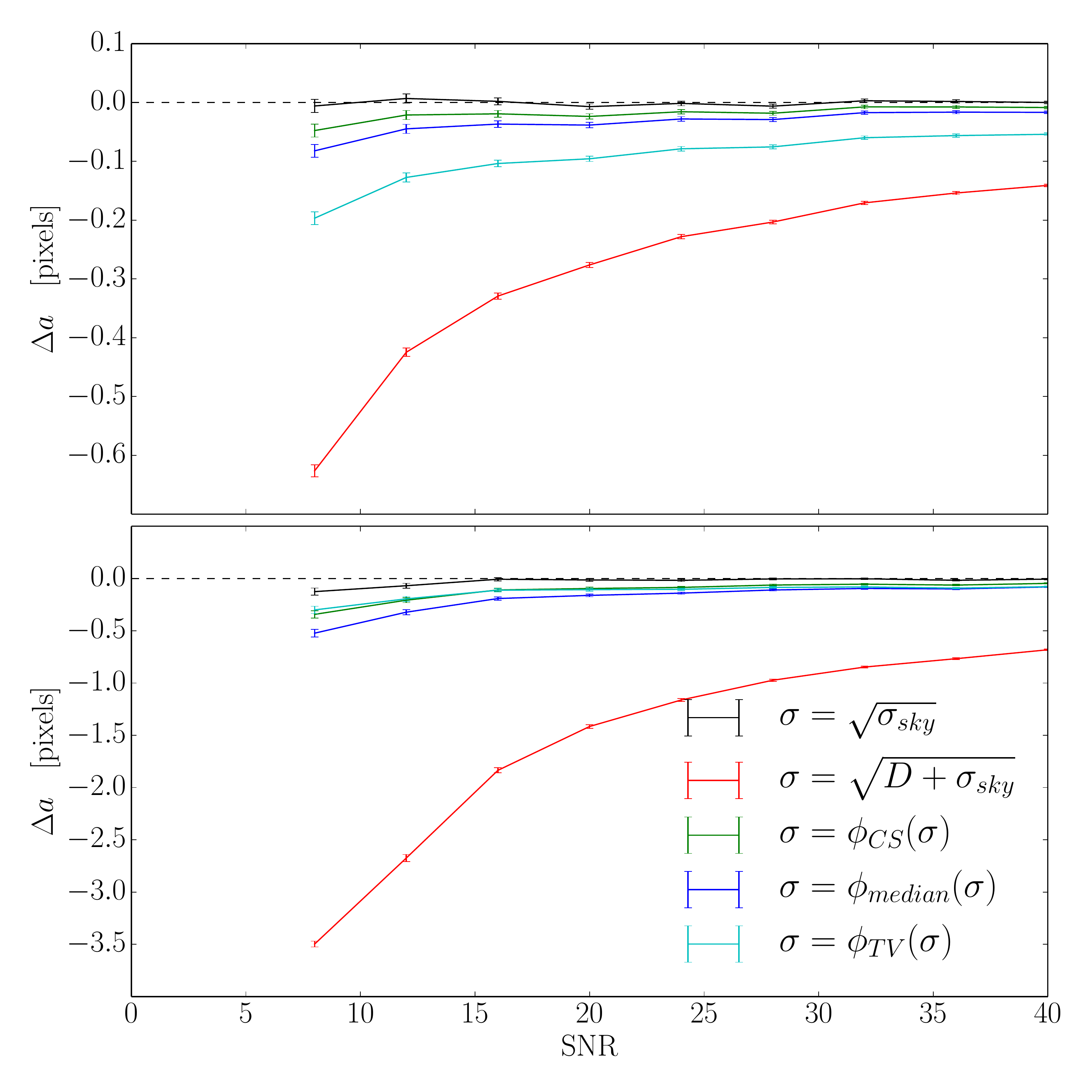}
\caption[]{Bias on the width of a circular Gaussian as a function of $S/N$. For each $S/N$ bin, the bias is measured using 1000 realizations of the simulation. {\it Top:} the fit without convolution by a PSF. {\it Bottom:} the fit with a PSF. Each curve shows the bias measured from the width of the simulated Gaussian for different ways of estimating the noise (see text).}
\label{fig:sig_dn}
\end{figure}

We then performed the Gaussian fit with and without PSF for different estimates of the noise map and compared the width of the fitted Gaussian to the true width. Figure~\ref{fig:sig_dn} summarizes our findings. The black line in the figure is obtained for Gaussian noise approximation and displays the noise bias found by \citet{Refregier2012}. This noise bias is best visible when the fit is made  with a PSF convolution. Although small, this affects weak-lensing experiments. If we now estimate the noise as in Eq.~\ref{shotnoise}, the bias we described above and illustrated in Fig.~\ref{fig:sigmabias} is prominent, and it is shown by the red curve. As this bias is due to the noise itself biasing the computation in Eq.~\ref{shotnoise}, we used a de-noised version of the data to estimate the noise amplitude map. This uses the de-noising scheme based on cycle-spinning described in Sect.~\ref{denoising}. We now estimated the noise amplitude map as
\begin{equation}
\boldsymbol{\sigma} = \sqrt{\phi_{2,3\sigma_{sky}}(\mathbf{D}) + \sigma_{sky}^2}.
\end{equation}

Using a de-noising filter before estimating the noise map results in a significant minimization of the bias, as shown by the green curve in Fig.~\ref{fig:sig_dn}. We also tested other de-noising schemes such as a median filter or a TV de-noising, which produced (except for TV de-noising) similar results. However, we always used cycle-spinning de-noising when we estimated the noise to carry out our deconvolutions because of its shape- and flux-preserving properties. \comment{malte}{cycle-spinning denoising is not the best name, but changing it would also require changing legend of Fig A.2}

\end{appendix}

\end{document}